%% file: The_power_of_Credit_Scoring_KPrzanowski.tex
\newcommand{\be}{\begin{equation}}
\newcommand{\ee}{\end{equation}}
\newcommand{\ba}{\begin{array}}
\newcommand{\ea}{\end{array}}
\newcommand{\beqn}{\begin{eqnarray}}
\newcommand{\eeqn}{\end{eqnarray}}

\newcommand{\zero}{\setcounter{equation}{0} 
}

\def\normalsize{
\setlength{\textheight}{23cm}
\setlength{\textwidth}{15cm}
\setlength{\topmargin}{-2.0cm}
\setlength{\hoffset}{0.2cm}
\setlength{\leftmargin}{-0.5cm}
\setlength{\rightmargin}{2.0cm}}

\documentclass[12pt]{article}
\usepackage[dvips]{graphicx}
\usepackage{natbib}
\DeclareGraphicsExtensions{.pdf,.png,.jpg,.eps}

\normalsize

\begin{document}
\title{Credit acceptance process strategy case studies - the power of Credit Scoring}
\date{}
\author{Karol Przanowski\\
Warsaw School of Economics –- SGH \\
Institute of Statistics and Demography \\
ul.Madalinskiego 6/8,\\
02-513 Warszawa  \\
email: \texttt{kprzan@sgh.waw.pl} \\
url: \texttt{http://kprzan.w.interia.pl}
}

\maketitle
\begin{abstract}
The paper is aware of the importance of certain figures that are essential to an understanding of Credit Scoring models
 in credit acceptance process optimization, namely if the power of discrimination measured by Gini value is increased by 5\% then the profit of the process can be increased monthly by about 1~500~kPLN (300~kGBP, 500~kUSD,  350~kEUR).
Simple business models of credit loans are also presented: acquisition - installment loan (low price) and cross-sell - cash loans (high price). 
Scoring models are used to optimize process, to become profitable. 
Various acceptance strategies with different cut-offs are presented, some are profitable and some are not.
Moreover, in a time of prosperity some are preferable whilst the inverse is true during a period of high risk or crisis.
To optimize the process four models are employed: three risk models, to predict the probability of default and one typical propensity model to predict the probability of response. It is a simple but very important example of the Customer Lifetime Value (CLTV or CLV) model business, where risk and response models are working together to become a profitable process.
\end{abstract}

\centerline{{\bf Key words:} credit scoring, crisis analysis, banking data generator, retail}
\centerline{portfolio, scorecard building, predictive modeling, credit acceptance process.}



\section{Introduction}
\zero

\input{wstep.tex}

\input{wykorz.tex}

\input{model_biznesowy.tex}

\section{Appendix}
\zero

\input{abt.tex}

\input{dokum.tex}



\newcommand{\byauthors}[1]{#1 }
\newcommand{\journal}[1]{ #1 }
\newcommand{\reftitle}[1]{{\it #1} }
\newcommand{\volumin}[1]{{\bf #1} }
\newcommand{\eref}{.}
\newcommand{\refyear}[1]{(#1), }




\end{document}

%% file: wstep.tex

In this paper some typical predictive models called Credit Scoring models or scorecards~\citep{crookbook, tolkit} are considered. 
These models are created based on logistic regression, especially the most known WoE approach~\citep{sasbook}. Their construction is very simple and useful in interpretation, so they have become the best tools in the optimization of processes in many financial institutions. For example, they are used in banking~\citep{huangimportant} to optimize credit acceptance processes and in PD models (probability of default) in regulator recommendations Basel II/III to calculate RWA (Risk Weighted Assets)~\citep{baselbis}.

Credit Scoring models are examples of statistical predictive models for forecasting some events based on collected and available data history. The best way to gauge their value is to wait and test them on real data, or, in other words, to compare the predicted values with observed. 
Unfortunately, this can take a considerable amount of time. To test the whole process, including the steps necessary in legal collection, can take between 5 and 10 years.

Credit Scoring research as a typical applied statistics field is fully connected with data science and with studies on real data taken from business banking processes. 
Legal constraints and lack of perspective thinking among data owners
can result in very real problems connected to access to real data
 and almost totally blocks any correctly led research. 
A change in thinking has occurred in biostatistics where access to real data is now possible, profiting both data scientists and data owners.
Even sometimes data are taken from the reality, however, in most cases they are insufficient to provide the complex analyses which are expected.

In the very interesting new paper presented at the conference {\it Credit Scoring and Credit Control XIII} in Edinburgh~\citep{Lessmanna_baza} all the available data used in the last ten years are presented. Many of them are used only once in order to highlight particular results. Only a few of them can be accessed by the public, but only one dataset has more than twenty variables and another with more than 150k observations. Based on this information it becomes clear that there is a need for new simulated data for the Credit Scoring analyses described by~\citep{kennedy}.

Let a set of simulated data be considered. All the rules used for data creating are known. Even if the Monte Carlo simulation, which is based on random numbers, is used repeatedly, all the final numbers are totally deterministic and can be repeated once again with the same number sequences. The following conclusion can, as a result, be reached: that research using this kind of data is ultimately futile, because all the previously held rules confirm only the method of data simulating. It is not true, because the method of data creating are completely different than scoring techniques, so it is nontrivial problem to find out description of the data based on scoring models. Moreover simulated data have many properties not planed and quite interesting. Complexity of the process is too hard to imagine and explain all unexpected behaviour, so also the author of the data can be surprised. There is needed a deeply study to point out total secret structures of new simulated data.

All the above mentioned arguments lead us to focus on simulated data. Furthermore, they also go so far as to suggest a change to a paradigm in applied statistics. Namely there is no requirement to always commence scientific research from the point of access to data. Maybe the question should be formulated as: what data are needed to ensure control of the process being studied in order to predict the future indicators?
It can be extremely risky to believe that real data can be sufficient. Observed data does not show latent variables. Simply because it has been observed does not mean that it is possible to explain. It should always be remembered that there is a necessity to focus on the hidden or latent information. It is for this reason that simulated data can be very useful, because invisible numbers, for example, risk measures calculated on rejected applications that are unobservable in reality, can be presented.  This is a very valuable rule created by Total Quality Management (TQM) suggesting that decisions should be made on both: visible and invisible numbers.

The need of creating simulated data can also be described in the following way: in the typical credit acceptance process reports are made known such as decline reasons, vintage, flow-rates, profile customer evaluation in the time, segmentations etc. All these reports are the results of observed information. The question what part do the hidden processes play in these results, could be considered.

%% file: wykorz.tex
\section{Simulated data creating, instalment loans case}
\zero

Consumer finance data generator in the first form is described by~\citep{karol_gen1}. It is dataset dedicated to only one product: an instalment loan. Every client has only one loan. All variables therefore are based only on one account history. The Markov chain and fixed transition matrix with calculated migration coefficients between states defined as a number of delinquent instalments per month is used. For every month on every account dedicated scoring connected with all available history up to current month is calculated. Every succeeding month of data is calculated based on the mentioned matrix and scorings. If an account score is low then in the next month the account under consideration has some due instalments dependent on the scoring bracket, the score belongs to better brackets, then the account remains in the current state or goes into a non-delinquent state.

Dataset contains 2~694~377 rows and 56 columns.

\section{Credit acceptance process profitability, predictive power impact}
\zero

It is obvious that Credit Scoring models are used in banking processes in the optimization purpose. It cannot be questioned, but up to now there have been no direct numbers about the usefulness of scorings presented in references, or about the profit values dependent on the scoring discrimination powers.
This is probably due to the secret know how of enterprises. This is why random simulated data can be useful. The case under consideration is not connected with any secret bank indicators, but, on the other hand, some estimated, relatively realistic numbers to imagine the power of scoring models and the main, key success factor in the banking business.

The source simulated data are specially changed to have a global risk, on all rows, at the level of 47\%. Following this, a few scoring models are constructed with different discrimination powers measured by Gini statistics~\citep{sasbook}.

It is not possible to present detailed Profit\&Loss (P\&L) report without knowledge about certain specific factors connected to the business field and market, but it is sufficient to focus on the main dimensions: incomes coming from interest rates and provisions; and losses calculated by Basel II/III recommendations as an expected loss.

The following definitions are necessary to understand profit notion:
${\rm APR}$ - annual percentage rate for credit loans, $r={{\rm APR}\over 12}$, $p$ - provision for credit granting paid at the beginning,
$A_i$ - loan amount, $N_i$ - number of instalments, where $i$ is index, account ID. 
Based on current Basel recommendations expected loss (EL) is defined as a multiplication of three factors: probability of default (PD), loss given default (LGD) and exposure at default (EAD).
Without any special argumentations, using a conservative approach it can be assumed that: LGD=50\% and EAD is the loan amount. 
Based on historical data, where default events are available, PD can be transformed from expected value into an observed one, namely when a default event is present (it can be written as ${\rm default}_{12}={\rm BAD}$) then PD=100\% and where the opposite is the case PD=0\%. In this case the observed loss $L_i$ is calculated. Incomes $I_i$ including provisions are calculated based on compound interest (geometric series). 
For every $i$ - account:

\begin{displaymath}
L_{i} = \left\{ \begin{array}{ll}
50\% A_i, & \textrm{\rm when default}_{12}={\rm BAD}, \\
0, & \textrm{\rm when default}_{12} \ne {\rm BAD}.
\end{array} \right.
\end{displaymath}

\begin{displaymath}
I_{i} = \left\{ \begin{array}{ll}
A_i p, & \textrm{\rm when default}_{12}={\rm BAD}, \\
A_i \left(
N_i r
{( 1+r)^{N_i} \over (1+r)^{N_i} -1} + (p-1) \right)
, & \textrm{\rm when default}_{12} \ne {\rm BAD}.
\end{array} \right.
\end{displaymath}

The total profit $P$ can be calculated as follows:
\beqn
P = & \sum_{i} I_i - L_i.
\label{profit}
\eeqn

\begin{figure}
\caption{Profit curves.}
\label{wszystkie_krzywe}
\vskip0.5cm
\begin{center}
\includegraphics[angle=0, width=0.8\textwidth]{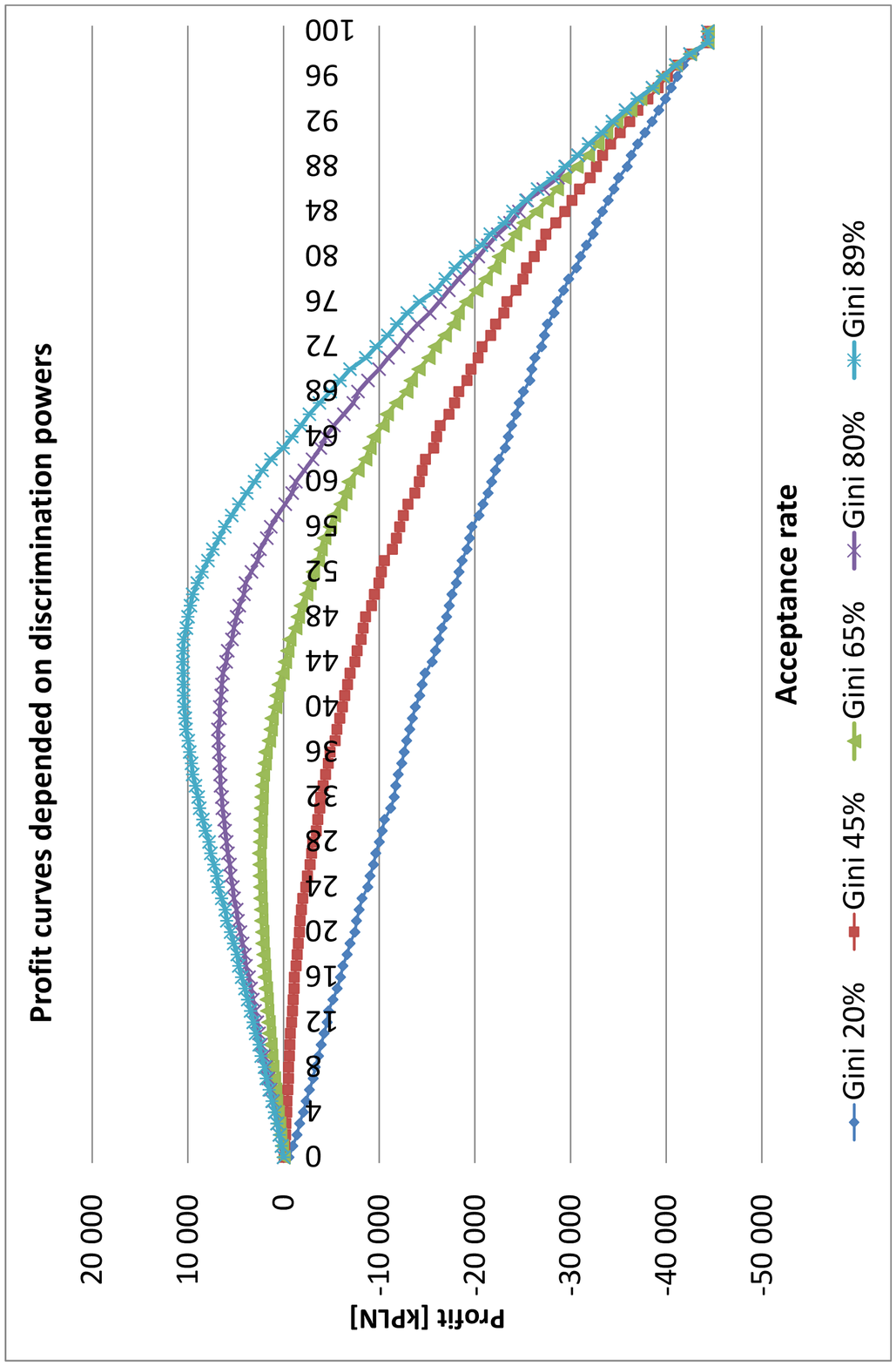}
\end{center}
\end{figure}

For every scoring model, with different predictive powers, all the applications can be sorted using score values from the worst, with the highest risk, to the best, with the lowest risk. Based on the cut-off parameter any profit can be calculated on an accepted part and acceptance rate. Repeating that procedure many times for all possible cut-offs and all scoring models defined profit curves are constructed. These are presented in figure~\ref{wszystkie_krzywe}. Note that all numbers on pictures are presented in PLN, but some important indicators are also recalculated in GBP, USD and EUR.
Some of them, for Gini this is usually lower than 50\%, never produce profits, for any acceptance rate the total profit is always negative.
Based on this, any scoring models with a low power or a low power of all acceptance rules, is not possible to manage the business and be profitable. Moreover, for a scenario when all the accepted applications are present, the profit is likely to be negative: a level of -44,5~mPLN (-9~mGBP, -15~mUSD, -10~mEUR).

\begin{figure}
\caption{The three best profit curves.}
\label{najlepsze_krzywe}
\vskip0.5cm
\begin{center}
\includegraphics[angle=0, width=0.8\textwidth]{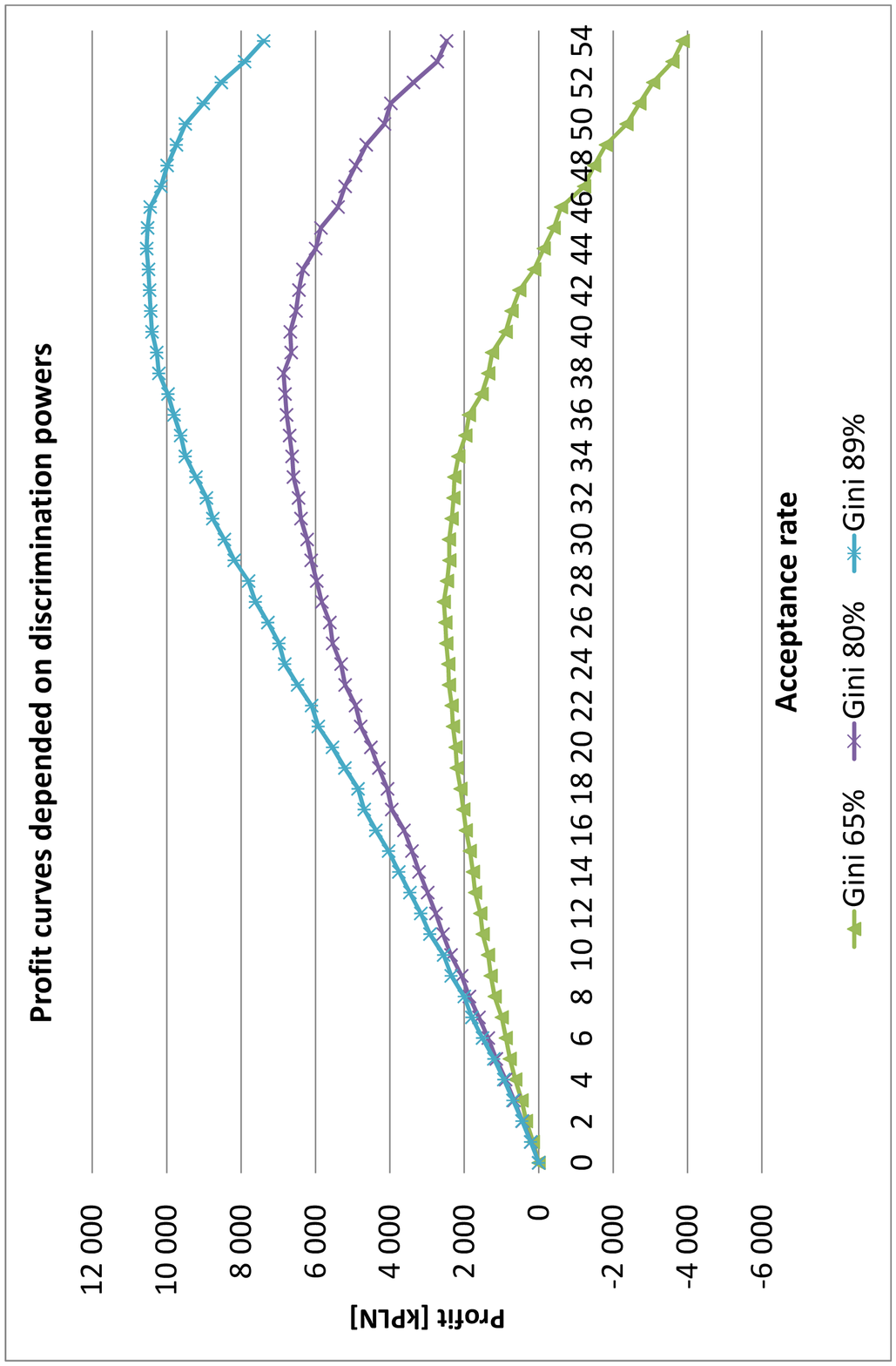}
\end{center}
\end{figure}

The best three curves, with possible profitable scenarios, are presented on figure~\ref{najlepsze_krzywe}. 
Scoring models with a power greater than about 50\% are able to identify profitable subsets of applications. 
The better the power, the greater the acceptance rate and profit. In the case of a scoring model with a power of 89\%, which is too high to be realistically entertained, can be accepted about 44\% of all applications resulting the total profit 10,5~mPLN (2~mGBP, 3.5~mUSD, 2.5~mEUR). 
These numbers are essential to understand the usage of scoring models. Dependent on the quality of the scoring model and its predictive power, a bank may lose or gain millions of currency units. Scoring is the key success factor in increasing capital for a company.

Profitability connected with correct usage of scoring models can also be presented in the following way. A calculation of the shift of profit and acceptance rate in the credit acceptance process when the increase of predictive power is equalled to 5\%, can be seen in table~\ref{moc_zysk1}.
If Gini is increased by 5\%, then the profit of the process can be 
increased monthly by about 1~500~kPLN (300~kGBP, 500~kUSD, 350~kEUR) the acceptance rate can be increased by 3,5\%. 
Alternatively, when an increase of acceptance rate is unnecessary, then a bank may save money only through the use of the better scoring model. Namely, with an acceptance rate set at 20\%, losses of approximately 900~kPLN (180~kGBP, 300~kUSD, 210~kEUR) can be saved. In the case of 40\% approximately 1~500~kPLN (300~kGBP, 500~kUSD, 350~kEUR) can be saved monthly.

The large amounts of potential profits or saved losses that are here presented may help persuade banks and other financial institutions to retain analytical teams in their employment. They may also help simulate the search for better models and a recognition of the need to test any new model that appears. Finally, they acknowledge the importance of necessity of champion challengers and parallels acceptance scenarios.

The general message seems to indicate that providing the figures quoted above are correct, then any changes in the acceptance rate are influenced by Reject Inference~\citep{huangimportant}. The inability to make correct risk estimations on rejected applications results in a substantially large bias in the profit estimation, so it is by no means straightforward to manage
 the credit acceptance process. Section~\ref{scenariusze} deals with it in detail.

\begin{table}
\begin{center}
\caption{
Shifts for finance indicators dependent on predictive power change.}
\label{moc_zysk1}
\vskip0.5cm
{\scriptsize

\begin{tabular}{ c | c }
\hline
Indicator & Value \\
\hline

Number of credit applications per month & 50~000 \\
Average loan amount & 5~000~PLN (1~kGBP, 1.6~kUSD, 1.1~kEUR) \\
Average number of instalments & 36 months \\
Annual percentage rate & 12\% \\
Provision for loan granting & 6\% \\
Global portfolio risk & 47\% \\
Increase of predictive power & 5\% \\
Increase of acceptance rate & 3,5\% \\
Increase of monthly profit & 1~500~kPLN (300~kGBP, 500~kUSD, 350~kEUR) \\
Decrease of monthly loss (AR=20\%) & 900~kPLN (180~kGBP, 300~kUSD, 210~kEUR) \\
Decrease of monthly loss (AR=40\%) & 1~500~kPLN (300~kGBP, 500~kUSD, 350~kEUR) 
\\ \hline
\end{tabular}
}
\end{center}
\end{table}

%% file: model_biznesowy.tex
\section{Business model: acquisition and cross-sell}
\zero

The last crisis (between 2008 and 2009) was a period when risk on some credit products, especially cash and revolving loans, dramatically increased. Consequently, many banks decided to decrease production, e.g. acceptance rates, to stabilize risk, trying to not exceed loss values and so become non-profitable.
Currently, banks, in the hope that the crisis has passed, have observed very low risk in their portfolios. There are some cases where this is actually lower than prior to the crisis. Consequently, many have taken the decision to increase acceptance rates.
In the current climate this has led to a struggle to attract customers.
Because the observed risk is presently relatively low, some customers segments are not profitable. 
In the case where a customer does not want to take a loan credit,
the bank is attempting to attract them by minimizing the price (e.g. APR or provision). This sometimes results in no profit gain.
A customer desperate for a loan is, in most cases, an extremely risky proposition. The fight is on to find customers with the right level of risk. In order to retain business, banks have to find a balance between
 profitable and overly risky customers.

The above results in the development or improvement of two-stage business models: the low and attractive model or the null price acquisition model, when a customer is started to establish an emotional relation with our bank, and then provide him many expensive repeat business products, cross-sell.

One of the most well-known two-stage business models in banking is: instalment loan as an acquisition and cash loan as a cross-sell. A Customer purchasing a TV-plasma in a store in small instalments is usually quite satisfied with the arrangement.
 During the loan cycle the customer gets many cross-sell contacts
or offers of cash loans.
 Some customers decide to purchase a further product and a subset of them are transformed into regular cash loan taking customers – a very profitable segment. Even if the business model is known it is not easy to manage it and to maximize a profits. This is the best example of where the importance of Credit Scoring Models can be illustrated.

Simulated data are also very useful in cut-off calculations. It is incorrect thinking to separate model building from its implementation. These two steps are always connected and Credit Scoring research cannot be focused only on various building techniques studies. When we want to build good models we need to test them in real production. Sometimes production results are quite different from those that are expected and an in-depth analysis of this difference is essential in the model building process.

It is an observed fact that future risk and instalment payments are correlated with available historical data on the customer. 
We can say that the customer's current ability to pay the next set of instalments is largely dependent upon their previous performance of repayments as well as upon their current financial, employment, home and domestic situation.
The customer also has their own priorities in loan repayments; some are paid regularly and some are paid before the due date; sometimes repayments are not made. These priorities are in some part connected with banks processes and collaterals, but the simplest way is to assume that any loan that is taken out alongside another loan will only serve to increase risk. Every customer has an ideal number of loans that they are able to successfully manage, but if the bank allows them to exceed this number, the customer may well default on repayments.

Let us consider one customer with applications for more than one loan. A bank may choose to grant all of them or only a part. 
As mentioned previously, the repayment of any successive loan is dependent on the history of previous repayments, but on only accepted and financed loans. It is not possible to consider all scenarios and to build model data in that case, but very simple solution is to assume that every application is somewhere accepted and then financed. 
When the customer applies for a loan for the first time he usually tries to get it from his favourite bank. If his application is rejected, he is likely to approach another bank, but if this application is turned down, then he may approach another type of financial institution or an individual.
We can assume that loans are always granted, but not always by the same bank.
 We can also use the basic economists thesis that expenses are not correlated with incomes. The same may be applied in the case of the granting of loans. The customer takes out a loan for his own individual reasons though these may not be connected to his affordability.

Let the basic assumptions of random data generator for Consumer Finance business model be formulated (acquisition = instalment loans and cross-sell = cash loans):
\begin{itemize}
\item The customer can only take out two types of loan: an instalment loan to purchase goods and a cash loan for any other purpose
\item Instalments loans are low risk and are not dependent on historical cash loans taken by the same customers
\item Cash loans are high risk and are dependent on the individual's repayment history: instalment and cash loans
\item The most risky loan for a customer is the last loan from any outstanding loans 
\item A cash loan can be granted in a particular month only when in the previous month a customer had some active accounts. In other words, every cash application is linked to publicity material dealing with offers that are only available to the bank's customers  
\item In any month there can only be one of two events: payment of several instalments or null payment, in databases information is collected regarding paid and due instalments
\item The distributions of characteristics are precisely defined using expert knowledge and are based on various random generators
\item If a customer has 7 due instalments on his account (180 past due days), then the account is closed with the status B (Bad) and history is not produced for any succeeding months
\item If all payments are made, then account is given a status C (closed correctly) and the history of that account is discontinued
\item Payments or non-payments are dependent on three factors: score value calculated on account and customer level (there are about 200 characteristics), transition matrix and macro-economic variable that changes the matrix over time.
\end{itemize}

All data are created on laptop Dell Latitude (1,67 GHz). 
Time of processing: 15 hours.

\begin{figure}
\caption{Instalment loan.}
\label{rap_ratalny}
\begin{center}
\includegraphics[angle=0, width=0.9\textwidth]{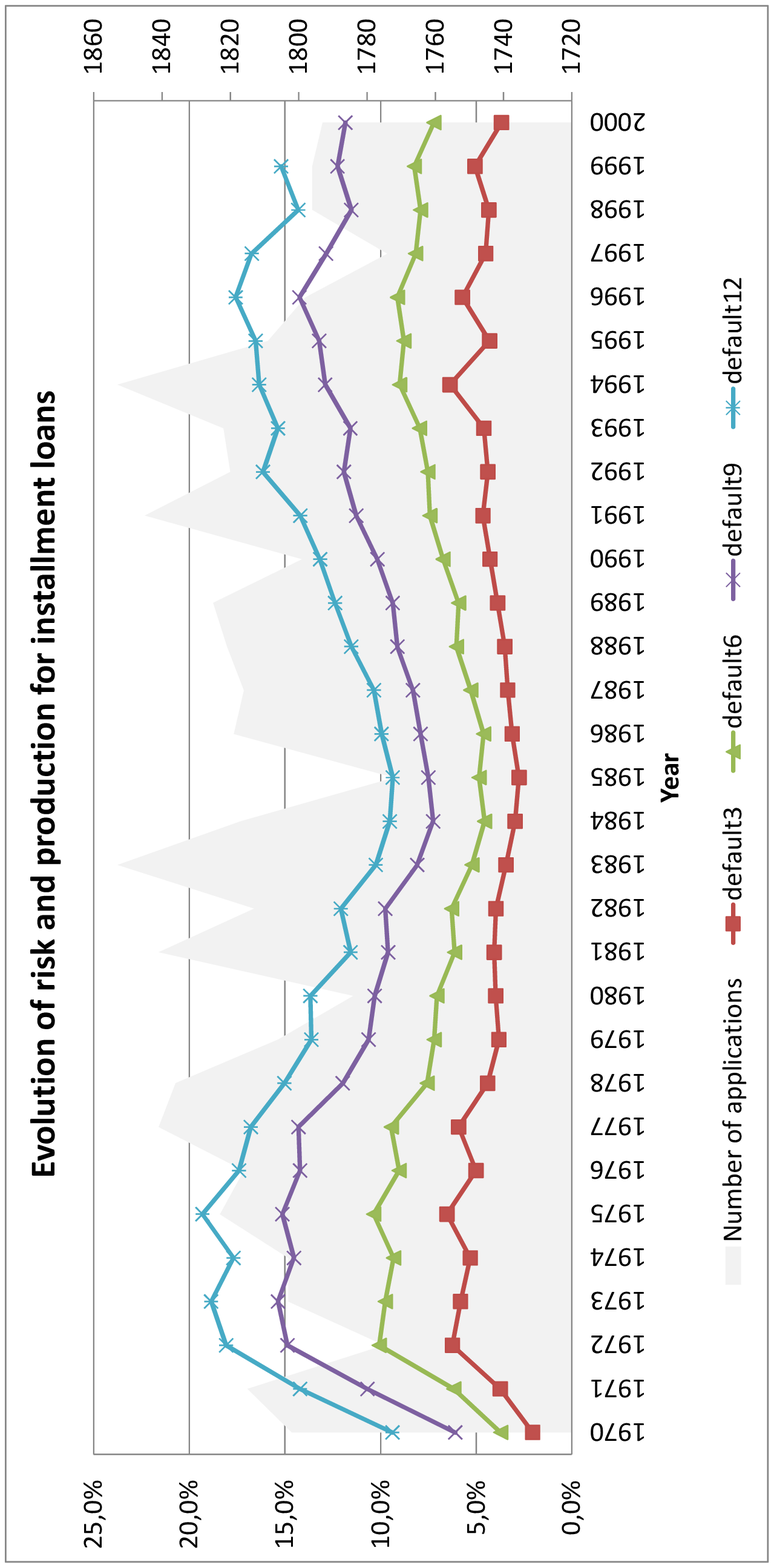}
\end{center}
\end{figure}

Datasets for instalment loans: Production dataset - 56~335 rows and 20 columns. Transaction dataset - 1~040~807 rows and 8 columns. All months are presented in figure~\ref{rap_ratalny}, where ${\rm default}_3$, ${\rm default}_6$, ${\rm default}_9$ and ${\rm default}_{12}$ means shares of accounts with 90+ past due days, three due instalments (in case ${\rm default}_3$ exceptionally 60+, two due instalments).

\begin{figure}
\caption{Cash loan.}
\label{rap_gotowka}
\begin{center}
\includegraphics[angle=0, width=0.9\textwidth]{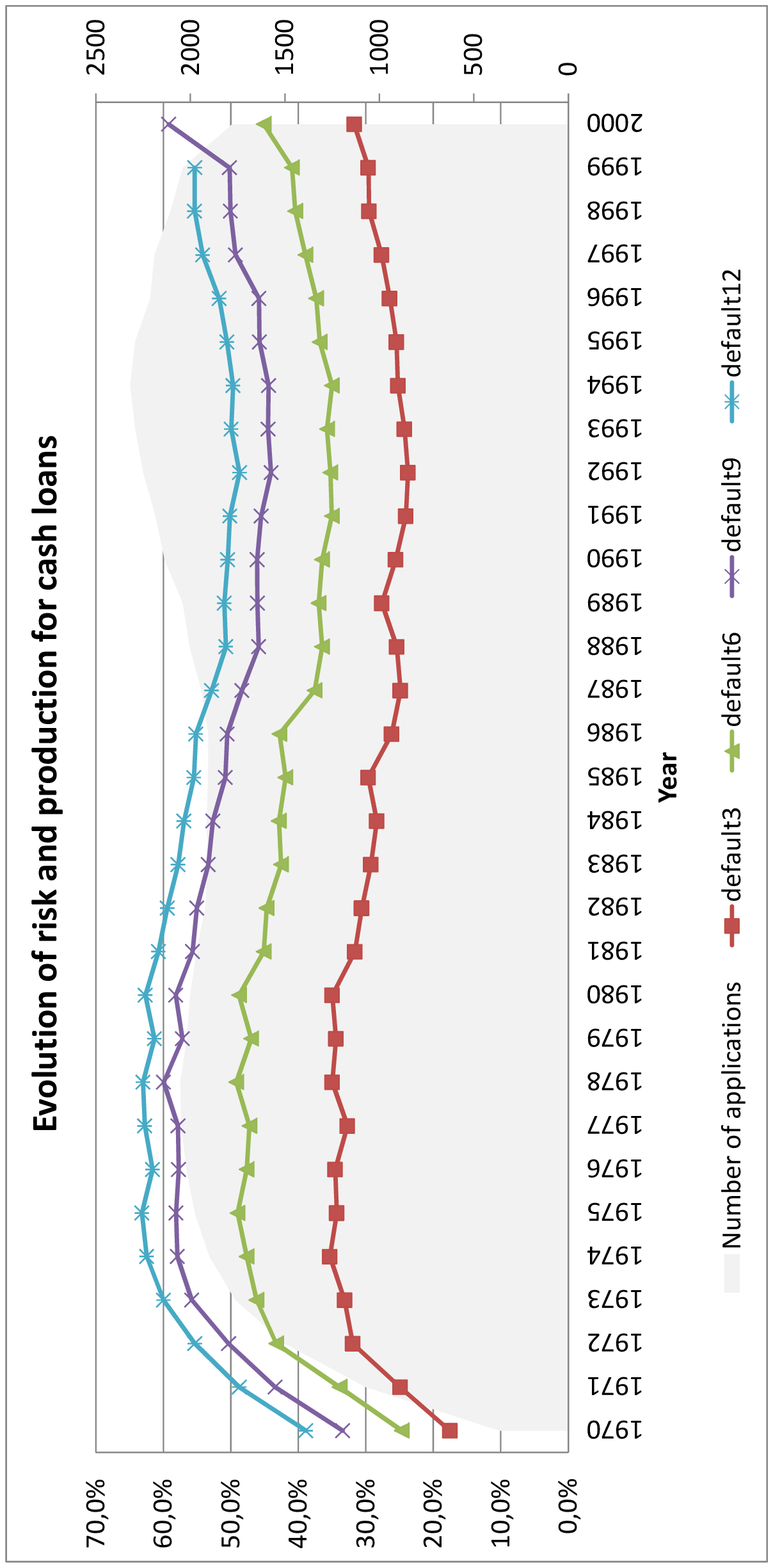}
\end{center}
\end{figure}

Datasets for instalment loans: Production dataset - 60~222 rows and 19 columns. Transaction dataset - 1~023~716 rows and 8 columns. All relevant months are presented in figure~\ref{rap_gotowka}. 
Additionally, the time period is divided into two periods: 1975-1987 -  modelling dataset, where all parameters are calculated and 1988-1998 for testing.

\begin{figure}
\caption{Stock months.}
\label{rap_stock}
\begin{center}
\includegraphics[angle=0, width=0.9\textwidth]{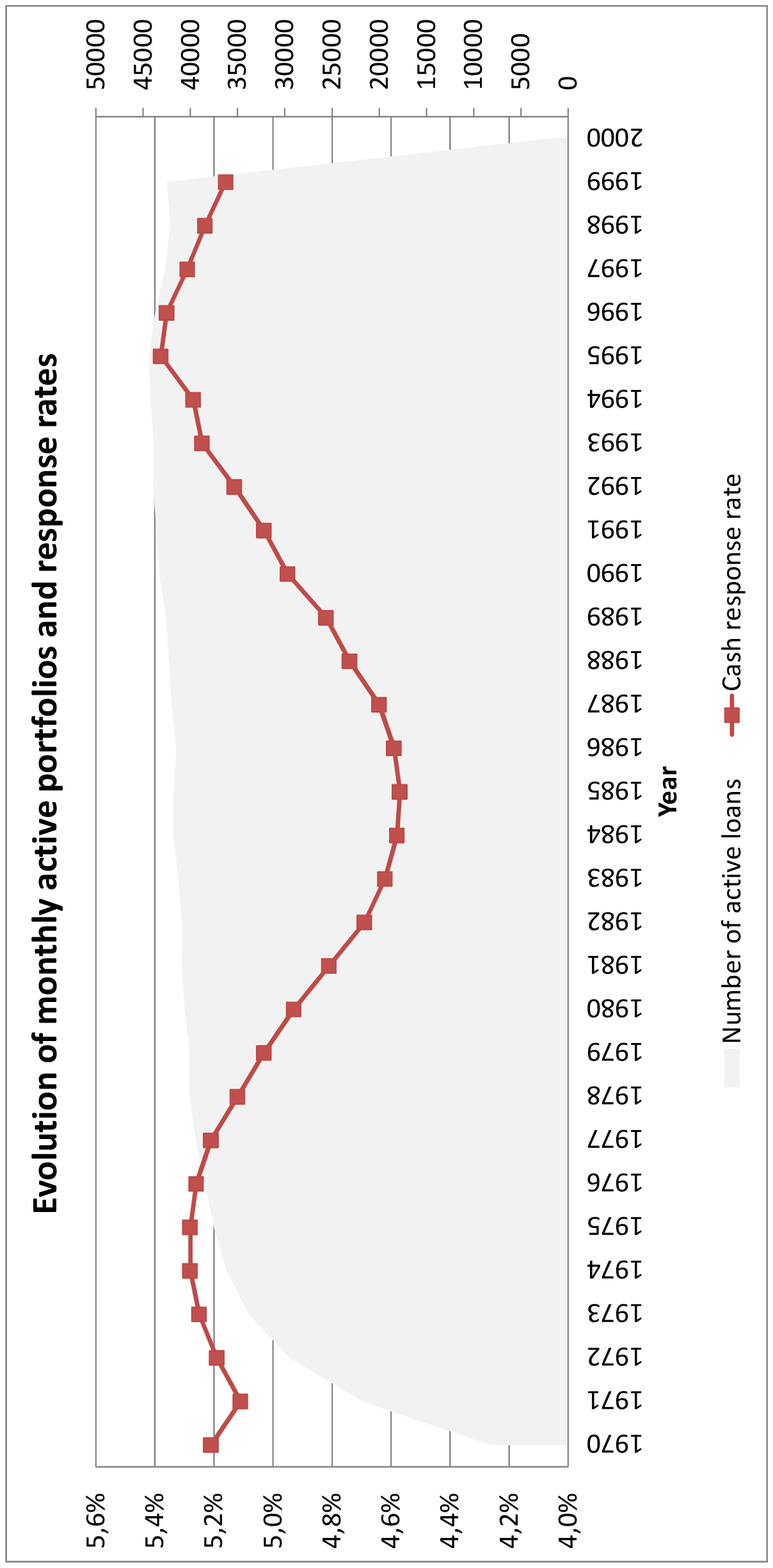}
\end{center}
\end{figure}

Stock months are presented in figure~\ref{rap_stock}. The response rate  equalled to about 5\% calculated as the number of total cash applications in the following month over the total number of active accounts in the potential market in the current month is also presented.

\section{Basic parameters calculation \label{silnik}}
\zero
All the simulated data, instalments and cash loans are available for a bank, allowing them the potential to expand on their role in the market. The customer has his own priority regarding repayments, some of their liabilities are prioritised; others are not. A bank can define a proper policy and set of acceptance rules to minimize any loss. Optimization is possible due to scoring models implemented in the decision engine, a specially dedicated IT tool for automatic processing. The bank can only optimize risk by its decisions: to accept or reject a particular applied request.
Acceptance implies inserting the whole generated history of processed application into bank's portfolio. It is assumed that all the account history is calculated before and is unchanging, so the issue is only to make a correct decision at the moment of the application based solely on any available customer history up to date of application. If a decision is negative then the bank does not have the history of that account in its portfolio.

The bank, depending upon its decision, has either a better or poorer knowledge of its customers. This poses an interesting question, which is better? To have available all information about a customer, about all his loan histories, but have a greater loss, or to have less information with lower a loss. Every rejected application of a particular customer results in a worse estimation of risk for future loans of that customer. The sample in the bank's portfolio of that customer is biased. The problem is called Reject Inference and today is described in many works~\citep{huangimportant,risaswp305,risas1,risas2,risas3,risas4,risas5,risas6}. The bank cannot avoid the above-mentioned problem, it can only minimize it or get more data from credit bureau companies. If the credit bureau in the country is managed properly and has all loans available in the market, then that information can be very useful in minimizing reject inference.

When bank knows more about his customers then it is able to estimate any potential risk in a better way, so it is able to provide its business a more stable and safer manner.

Four models are constructed: three risk models and one response model. All the models are built on the same sample time period 1976-1987: 
\begin{itemize}
\item PD Ins - PD model to predict probability of instalment ${\rm default}_{12}$.
\item PD Css - PD model to predict probability of cash ${\rm default}_{12}$.
\item Cross PD Css - PD model to predict probability of future cash ${\rm default}_{12}$ at the time of instalment application.
\item PR Css - response model to predict probability of future cash response at the time of instalment application.
\end{itemize}

Model documentations are presented in~\ref{dok_modeli}.

The main goal is to make the process profitable and to maximize profit. Simple factors are calculated: income - interest rate incomes and loss - expected loss based on known Basel formula: EL=PD*LGD*EAD see also formula~\ref{profit}. Parameters are set as follows:

The annual percentage rate (APR) for an instalment loan is $1\%$, APR for cash loans $18\%$. Average LGD values: $45\%$ for instalment and $55\%$ for cash loans. PD for EL formula is binary variable ${\rm default}_{12}$. To simplify the cross-sell process all cash loans are the same: loan amount  $5 000$PLN and the number of instalments - $24$.

Financial KPIs for the modelling period 1975-1987 are presented in table~\ref{sc0_all_fin}. Table~\ref{sc0_moc_fin} provides predictive the powers of built and used models, some powers are not realistic, especially for the response model, but that case represents a strategy of full acceptance in $100\%$, which is a fairly unreal scenario and is considered only due to random data. It is useful to study it, because we can observe a case without any reject inference.

\begin{table}
\begin{center}
\caption{Finance KPIs for global process, where all applications are accepted (period 1975-1987).}
\label{sc0_all_fin}
\vskip0.5cm

\begin{tabular}{ l | r | r | r }
\hline
KPI & Instalment & Cash & All \\
\hline
Profit & -7 824 395 & -31 627 311 & -39 451 706 \\
Income & 969 743 & 10 260 689 & 11 230 432 \\
Loss & 8 794 138 & 41 888 000 & 50 682 138 \\
\hline
\end{tabular}
\end{center}
\end{table}

\begin{table}
\begin{center}
\caption{Predictive powers (period 1975-1987).}
\label{sc0_moc_fin}
\vskip0.5cm

\begin{tabular}{ l | r }
\hline
Model & Gini \\
\hline
Cross PD Css & 74,01\% \\
PD Css & 74,21\% \\
PD Ins & 73,11\% \\
PR Css & 86,37\% \\
\hline
\end{tabular}
\end{center}
\end{table}

The average risk value of this process is 37,19\% and the average probability (PD) - 34,51\%, so the expected value is slightly underestimated. The global profit is negative about -40 mPLN. 
The chance of making a profit is understandably not a straightforward task. However, a solution can be found that is based upon the Customer
 Lifetime Value (CLTV, or CLV) modelling methods~\citep{sasltv,modelecrm}. It uses a relatively simple version that is based upon one response three risk models.

The goal of maximizing profit can be achieved by finding proper cut-offs for the above-mentioned four models. We should also be aware of the general idea of processing. All customers and their loans with all relevant history are collected in a database that we can call portfolio potential. Our decision engine can only accept or reject loans. If some applications are rejected then some missing information about our customers is present. That missing one has an effect on risk expected values, on score distributions and on the ABT variables distributions described in section~\ref{opis_abt}. So on the one hand our bank can have lower risk, because some applications are rejected, but on the other some important information about customers is lost. Moreover, if some loans are not accepted, there may be some other type of value missing present, namely when a cash loan is being taken out. If we do not have a customer with active loans in our portfolio, we are not able to send them cash offers, so we do not know, or rather the customer does not know, they may access to cash in our bank, so we end up losing that customer. That kind of missing information in acceptance strategies is indicated as 'not known customer'.

First we try to optimize all cash loans. Based on a simple profit curve a proper cut-off on PD Css probability values can be found, namely with an acceptance rate of 18,97\% we have the best profit = 1~591~633~PLN. 
In the decision engine we introduce the {\bf rule:} when $PD\_Css>27,24\%$ then reject.

Following CLTV methodology the sequence of cash loans should be considered which would then be analysed in a more efficient manner to discover the final profit.
This exercise is studied only in the case when the first product is an instalment loan. Also considered is the future cash loan for the same customer. Of course, not every customer taking out an instalment loan is taking cash later. Only some of decide to take cash, so some customers, especially those with only instalment loans, are rather non-profit. Our process should be focused on customers with a bigger chance for future cash loan, because only from that kind of customer we can earn money.

Based on models PD Ins and PR Css five (from 0 to 4) segments are created separately. The first, early discovered rule is also considered. For every combination of groups the global profit is calculated taking into account the currently applied instalment loan and future cash loan (see table~\ref{zysk_global}). Based on that table new rules are defined: 

\begin{itemize}
\item [{\bf rule:}] when $PD\_Ins>8,19\%$, then reject
\item [{\bf rule:}] when $8,19\%>=PD\_Ins>2,18\%$ and 
	$(PR\_Css<2,8\%$ or \\
$Cross\_PD\_Css>27,24\%)$, then reject.
\end{itemize}

It should be emphasized, that the last rule is based not only on risk parameters. It can be interpreted in the following way: if a customer takes only an instalment loan, then the cut-off can be set on a different level than for customers who take cash loans in the future.

All the defined rules should result in 1~686~684~PLN of global profit. If the last rule is omitted, that is with only one acceptance rule: {\bf $PD\_Ins \le 8,19\%$}, then the global profit will be 1~212~261~PLN. So we will lose about 470 kPLN, which will be about 30\% lower profit.

Unfortunately, the numbers and ideas presented are biased by Reject Inference. To be sure of the final numbers we need to run the acceptance process, calculate once again all the ABT variables, based on new decisions, and then we will be able to get the proper financial KPIs. This is the reason for the various strategy testing in section~\ref{scenariusze}.

The method presented for cut-offs calculation is only an example and can be treated as a nice exercise to learn the complex processes and correlations between many factors. In the real-life situation more correctly defined goals, boundaries, constraints or restrictions should be considered. Nevertheless, the technique is always the same; all the possible scenarios are considered. Only then are the KPIs calculated and the best solution is decided upon. But the problem of reject inference is always present and it is not easy to consider its impact in an appropriate manner.

\begin{table}
\begin{center}
\caption{Combinations of segments (groups) and their global profits (period 1975-1987).}
\label{zysk_global}
\vskip0.5cm
{\scriptsize

\begin{tabular}{ r | r | r | r | r | r | r | r }
\hline
GR PR & GR PD & Number & Global & Min & Max & Min & Max \\
Css & Ins & of Ins applications & profit & PR Css & PR Css & PD Ins & PD Ins \\
\hline
4 & 0 & 1 277 & 372~856 & 4,81\% & 96,61\% & 0,02\% & 2,18\% \\
4 & 1 & 581 & 96~096 & 4,81\% & 96,61\% & 2,25\% & 4,61\% \\
1 & 0 & 2 452 & 67~087 & 1,07\% & 1,07\% & 0,32\% & 2,18\% \\
3 & 0 & 907 & 46~685 & 2,80\% & 4,07\% & 0,07\% & 2,18\% \\
3 & 1 & 734 & 14~813 & 2,80\% & 4,07\% & 2,25\% & 4,61\% \\
3 & 2 & 307 & 12~985 & 2,80\% & 4,07\% & 4,76\% & 7,95\% \\
4 & 2 & 361 & 8~039 & 4,81\% & 96,25\% & 4,76\% & 7,95\% \\
\hline

3 & 3 & 446 & -1~283 & 2,80\% & 4,07\% & 8,19\% & 18,02\% \\
4 & 3 & 417 & -5~774 & 4,81\% & 95,57\% & 8,19\% & 18,02\% \\
1 & 1 & 3 570 & -82~886 & 1,07\% & 1,07\% & 2,25\% & 4,61\% \\
1 & 2 & 4 044 & -408~644 & 1,07\% & 1,07\% & 4,76\% & 7,95\% \\
3 & 4 & 726 & -946~937 & 2,80\% & 4,07\% & 18,50\% & 99,62\% \\
4 & 4 & 1 054 & -1~108~313 & 4,81\% & 96,25\% & 18,50\% & 99,83\% \\
1 & 3 & 3 883 & -1~270~930 & 1,07\% & 1,07\% & 8,19\% & 18,02\% \\
1 & 4 & 2 878 & -4~306~859 & 1,07\% & 1,07\% & 18,50\% & 97,00\% \\
\hline

\end{tabular}
}
\end{center}
\end{table}

\section{Various strategies study\label{scenariusze}}
\zero

To understand the complexity of the process four acceptance strategies are presented, see tables~\ref{Strategy1}, \ref{Strategy2}, \ref{Strategy3} and~\ref{Strategy4}.

The first strategy is connected to the cut-offs calculation approach mentioned in section~\ref{silnik}. Let we us remember that in the period 1975-1987 the expected total profit is 1~686~684~PLN. After running the process we arrive at 663~327~PLN. There is an error of approximately one million PLN. What is going on? Where is our money? The error is very significant and it is because of rejected applications, 30\% of share, and due to not known customer decline reasons, 50\% of share. On the other hand, we can content ourselves in the knowledge that our profit is still positive by about 700~kPLN, as opposed to the total acceptance strategy with a negative profit equalling -40~mPLN. Despite this small success the error is too significant to ignore, it persuades us to study further and broaden our minds in order to be aware that building models with large predictive powers is not enough to win an implementation step. Taking into account all the factors and all the steps is also very important.

To be very honest, nobody can guarantee a success in any case where the strategy is dramatically changed. Let we emphasize that the initial strategy is based on the fully acceptance process, so from a 100\% acceptance rate is switched to 26\% on instalment loans and 16\% on cash. Such a radical step plays a profound part in the distributions of our probabilities. To begin with, PD is at a level of 34,51\%. In the new process we have 28,87\%. Why do we have a difference? It is due to missing information about all of the customer's accounts. The new strategy accepts less risky applications, so the bank has information only about better loans, so the average PD should be lower, which causes an underestimation of risk regarding our customers.

In the inverse case, based on first strategy, if we want to increase acceptance, we will be in trouble because we will have incorrectly estimated risks parameters. 
Moreover, we will realise that the properties of the models are also changed, and the predictive powers are lower. Gini of model Cross PD Css from 74\% decreased up to 41\% on all and to 21\% on accepted segment. In reality, only the second value is observed, so we can discuss about the correctness of that model. Why does the model not work? It is really difficult to measure the value of that model after implementation. Probably it will be replaced by another one.

In section~\ref{silnik} the result is also shown in the case of one simple rule without any special rules based on response probabilities. In that case the expected profit is lower by about 470~kPLN. It falls by about 550~kPLN (see strategy 2, table~\ref{Strategy2}), so in estimations of differences we do not have any significant errors.

We can formulate a simple rule: reject inference is difficult to predict, it is proportional to the strength of the acceptance process change. If we change a process significantly, we can also expect significant error in predictions due to rejected applications.

Let we consider another approach. Let's start from a simple intuition strategy, where we do not accept applications with default events (more than three due instalments) during the last 12 months before the application date. The process is not profitable (see strategy 3, table~\ref{Strategy3}), but acceptance of cash loans is decreased to 45\%. The models still work in that case. Repeating the same idea as for cut-offs calculation for the first strategy, we can once again set new cut-offs, but in that case based on a different starting strategy. We have therefore created the next strategy (see strategy 4, table~\ref{Strategy4}).

The profit is 732~kPLN, with 9\% acceptance for cash and 26\% for instalment loans. We can discuss this strategy further, it is a better solution for the period 1975-1987, but acceptance of cash loans is very small, which is only right for such a period of high risk. Indeed, in the period 1988-1998 we have the invers case: the first strategy has a profit of 1,5~mPLN but the forth a little bit lower 1,3~mPLN.

All the exercises presented give us the opportunity to understand complex problems and to be aware of reject inference in practice.

The most important conclusions are: because we are able to create useful data for Credit Scoring analysis, we can make many strategies, build various models, test CLTV approaches and develop better estimations of reject inference.

\begin{table}
\begin{center}
\caption{Strategy 1.}
\label{Strategy1}
\vskip0.5cm
{\scriptsize

\begin{tabular}{ r | r | r | r }
\hline
Period & Income & Loss & Profit \\
\hline
1975-1987 & 3 407 745 & 2 744 418 & 663 327 \\
1988-1998 & 3 761 299 & 2 246 844 & 1 514 455 \\
\hline
\end{tabular}

\vskip0.3cm
\begin{tabular}{ l | l }
\hline
Rule & Description \\
\hline
PD\_Ins Cutoff & $PD\_Ins>8,19\%$ \\
PD\_Css Cutoff & $PD\_Css>27,24\%$  \\
Special for PD and PR & $8,19\%>=PD\_Ins>2,18\%$ and $(PR\_Css<2,8\%$ or $Cross\_PD\_Css>27,24\%)$ \\
\hline
\end{tabular}

\vskip0.3cm
\centerline{Cash loan}
\begin{tabular}{ r | r | r | r | r | r }
\hline
Rule & Number of applications & \% of applications & Loan amount & Risk & Profit \\
\hline
PD\_Css Cutoff & 8 436 & 32,97\% & 42 180 000 & 67,99\% & -13 098 591 \\
Not known customer & 12 999 & 50,80\% & 64 995 000 & 65,91\% & -19 171 357 \\
Accepted & 4 152 & 16,23\% & 20 760 000 & 22,35\% & 642 637 \\
All & 25 587 & 100,00\% & 127 935 000 & 59,53\% & -31 627 311 \\
\hline
\end{tabular}

\vskip0.3cm
\centerline{Instalment loan}
\begin{tabular}{ r | r | r | r | r | r }
\hline
Rule & Number of applications & \% of applications & Loan amount & Risk & Profit \\
\hline
PD\_Ins Cutoff & 9 289 & 39,30\% & 60 214 008 & 26,95\% & -7 339 423 \\
Special for PD and PR & 8 131 & 34,40\% & 31 340 808 & 5,37\% & -505 662 \\
Accepted & 6 217 & 26,30\% & 22 698 240 & 2,14\% & 20 690 \\
All & 23 637 & 100,00\% & 114 253 056 & 13,00\% & -7 824 395 \\
\hline
\end{tabular}

\vskip0.3cm
\centerline{Average parameter values}
\begin{tabular}{ r | r | r }
\hline
Parameter & Accepted & All \\
\hline
PD (combined PD Ins i PD Css) & 7,93\% & 28,87\% \\
PR Css & 17,15\% & 21,76\% \\
Cross PD Css & 21,71\% & 17,73\% \\
\hline
\end{tabular}

\vskip0.3cm
\centerline{Predictive power (Gini)}
\begin{tabular}{ r | r | r }
\hline
Model & Accepted & All \\
\hline
Cross PD Css & 21,34\% & 40,72\% \\
PD Css & 31,66\% & 53,28\% \\
PD Ins & 41,93\% & 68,58\% \\
PR Css & 72,56\% & 68,88\% \\
\hline
\end{tabular}

}
\end{center}
\end{table}

\begin{table}
\begin{center}
\caption{Strategy 2.}
\label{Strategy2}
\vskip0.5cm
{\scriptsize

\begin{tabular}{ r | r | r | r }
\hline
Period & Income & Loss & Profit \\
\hline
1975-1987 & 4 008 258 & 3 896 818 & 111 441 \\
1988-1998 & 4 539 328 & 3 829 634 & 709 694 \\
\hline
\end{tabular}

\vskip0.3cm
\begin{tabular}{ l | l }
\hline
Rule & Description \\
\hline
PD\_Ins\ Cutoff & $PD\_Ins>8,19\%$ \\
PD\_Css\ Cutoff & $PD\_Css>27,24\%$  \\
\hline
\end{tabular}

\vskip0.3cm
\centerline{Cash loan}
\begin{tabular}{ r | r | r | r | r | r }
\hline
Rule & Number of applications & \% of applications & Loan amount & Risk & Profit \\
\hline
PD\_Css Cutoff & 9 297 & 36,33\% & 46 485 000 & 67,84\% & -14 381 482 \\
Not known customer & 11 661 & 45,57\% & 58 305 000 & 67,34\% & -17 822 432 \\
Accepted & 4 629 & 18,09\% & 23 145 000 & 23,16\% & 576 604 \\
All & 25 587 & 100,00\% & 127 935 000 & 59,53\% & -31 627 311 \\
\hline
\end{tabular}

\vskip0.3cm
\centerline{Instalment loan}
\begin{tabular}{ r | r | r | r | r | r }
\hline
Rule & Number of applications & \% of applications & Loan amount & Risk & Profit \\
\hline
PD\_Ins Cutoff & 9 325 & 39,45\% & 60 221 856 & 26,98\% & -7 359 232 \\
Accepted & 14 312 & 60,55\% & 54 031 200 & 3,89\% & -465 163 \\
All & 23 637 & 100,00\% & 114 253 056 & 13,00\% & -7 824 395 \\
\hline
\end{tabular}

\vskip0.3cm
\centerline{Average parameter values}
\begin{tabular}{ r | r | r }
\hline
Parameter & Accepted & All \\
\hline
PD (combined PD Ins i PD Css) & 6,82\% & 29,05\% \\
PR Css & 12,79\% & 22,89\% \\
Cross PD Css & 17,62\% & 18,34\% \\
\hline
\end{tabular}

\vskip0.3cm
\centerline{Predictive power (Gini)}
\begin{tabular}{ r | r | r }
\hline
Model & Accepted & All \\
\hline
Cross PD Css & 19,39\% & 39,86\% \\
PD Css & 31,23\% & 55,05\% \\
PD Ins & 41,73\% & 69,04\% \\
PR Css & 80,56\% & 64,40\% \\
\hline
\end{tabular}

}
\end{center}
\end{table}

\begin{table}
\begin{center}
\caption{Strategy 3.}
\label{Strategy3}
\vskip0.5cm
{\scriptsize

\begin{tabular}{ r | r | r | r }
\hline
Period & Income & Loss & Profit \\
\hline
1975-1987 & 7 496 614 & 21 801 230 & -14 304 616 \\
1988-1998 & 7 881 992 & 18 510 342 & -10 628 350 \\
\hline
\end{tabular}

\vskip0.3cm
\begin{tabular}{ l | l }
\hline
Rule & Description \\
\hline
Bad customer & $agr12\_Max\_CMaxA\_Due>3$ \\
\hline
\end{tabular}

\vskip0.3cm
\centerline{Cash loan}
\begin{tabular}{ r | r | r | r | r | r }
\hline
Rule & Number of applications & \% of applications & Loan amount & Risk & Profit \\
\hline
Bad customer & 7 114 & 27,80\% & 35 570 000 & 79,83\% & -14 195 320 \\
Not known customer & 7 036 & 27,50\% & 35 180 000 & 67,04\% & -10 673 871 \\
Accepted & 11 437 & 44,70\% & 57 185 000 & 42,28\% & -6 758 120 \\
All & 25 587 & 100,00\% & 127 935 000 & 59,53\% & -31 627 311 \\
\hline
\end{tabular}

\vskip0.3cm
\centerline{Instalment loan}
\begin{tabular}{ r | r | r | r | r | r }
\hline
Rule & Number of applications & \% of applications & Loan amount & Risk & Profit \\
\hline
Bad customer & 483 & 2,04\% & 2 047 188 & 27,74\% & -277 899 \\
Accepted & 23 154 & 97,96\% & 112 205 868 & 12,69\% & -7 546 496 \\
All & 23 637 & 100,00\% & 114 253 056 & 13,00\% & -7 824 395 \\
\hline
\end{tabular}

\vskip0.3cm
\centerline{Average parameter values}
\begin{tabular}{ r | r | r }
\hline
Parameter & Accepted & All \\
\hline
PD (combined PD Ins i PD Css) & 21,81\% & 32,70\% \\
PR Css & 21,79\% & 28,83\% \\
Cross PD Css & 43,09\% & 24,48\% \\
\hline
\end{tabular}

\vskip0.3cm
\centerline{Predictive power (Gini)}
\begin{tabular}{ r | r | r }
\hline
Model & Accepted & All \\
\hline
Cross PD Css & 64,83\% & 63,59\% \\
PD Css & 63,67\% & 64,82\% \\
PD Ins & 71,94\% & 72,56\% \\
PR Css & 79,96\% & 64,72\% \\
\hline
\end{tabular}

}
\end{center}
\end{table}

\begin{table}
\begin{center}
\caption{Strategy 4.}
\label{Strategy4}
\vskip0.5cm
{\scriptsize

\begin{tabular}{ r | r | r | r }
\hline
Period & Income & Loss & Profit \\
\hline
1975-1987 & 2 010 242 & 1 278 361 & 731 882 \\
1988-1998 & 2 452 716 & 1 134 729 & 1 317 986 \\
\hline
\end{tabular}

\vskip0.3cm
\begin{tabular}{ l | l }
\hline
Rule & Description \\
\hline
Bad customer & $agr12\_Max\_CMaxA\_Due>3$ \\
PD\_Ins\ Cutoff & $PD\_Ins>7,95\%$ \\
PD\_Css\ Cutoff & $PD\_Css>19,13\%$  \\
Special for PD and PR & $7,95\%>=PD\_Ins>2,8\%$ and $(PR\_Css<2,8\% lub Cross\_PD\_Css>19,13\%)$ \\
\hline
\end{tabular}

\vskip0.3cm
\centerline{Cash loan}
\begin{tabular}{ r | r | r | r | r | r }
\hline
Rule & Number of applications & \% of applications & Loan amount & Risk & Profit \\
\hline
Bad customer & 2 253 & 8,81\% & 11 265 000 & 74,26\% & -4 026 033 \\
PD\_Css Cutoff & 5 375 & 21,01\% & 26 875 000 & 53,66\% & -5 462 687 \\
Not known customer & 15 739 & 61,51\% & 78 695 000 & 65,29\% & -22 845 756 \\
Accepted & 2 220 & 8,68\% & 11 100 000 & 17,97\% & 707 165 \\
All & 25 587 & 100,00\% & 127 935 000 & 59,53\% & -31 627 311 \\
\hline
\end{tabular}

\vskip0.3cm
\centerline{Instalment loan}
\begin{tabular}{ r | r | r | r | r | r }
\hline
Rule & Number of applications & \% of applications & Loan amount & Risk & Profit \\
\hline
Bad customer & 209 & 0,88\% & 891 720 & 27,75\% & -121 550 \\
PD\_Ins Cutoff & 9 253 & 39,15\% & 60 130 704 & 26,46\% & -7 208 030 \\
Special for PD and PR & 8 029 & 33,97\% & 31 118 232 & 5,49\% & -519 531 \\
Accepted & 6 146 & 26,00\% & 22 112 400 & 2,05\% & 24 717 \\
All & 23 637 & 100,00\% & 114 253 056 & 13,00\% & -7 824 395 \\
\hline
\end{tabular}

\vskip0.3cm
\centerline{Average parameter values}
\begin{tabular}{ r | r | r }
\hline
Parameter & Accepted & All \\
\hline
PD (combined PD Ins i PD Css) & 4,24\% & 25,17\% \\
PR Css & 11,37\% & 15,68\% \\
Cross PD Css & 17,02\% & 14,61\% \\
\hline
\end{tabular}

\vskip0.3cm
\centerline{Predictive power (Gini)}
\begin{tabular}{ r | r | r }
\hline
Model & Accepted & All \\
\hline
Cross PD Css & 3,23\% & 19,19\% \\
PD Css & 33,15\% & 47,81\% \\
PD Ins & 36,79\% & 67,67\% \\
PR Css & 70,59\% & 64,89\% \\
\hline
\end{tabular}

}
\end{center}
\end{table}

\section{Conclusions}
\zero

Credit Scoring models in credit acceptance process are the best tools for optimization and maximisation of outcome profit.
 Profitability connected with correct usage of scoring models can be presented in the following way. If predictive power measured in Gini statistic is increased by 5\% then the profit of the process can be 
increased monthly by about 1~500~kPLN (300~kGBP, 500~kUSD, 350~kEUR) and acceptance rate by 3,5\%. In other words, predictive power of scoring models is an important factor to earn millions of currency units monthly.

Risk estimation is very difficult to predict of a significant change of the acceptance process. In that case impact of rejected applications, called Reject Inference is difficult to predict and all estimations are made on biased sample.

Usage of simulated data in credit acceptance process research can reveal some hidden, invisible numbers, like the risk value that is present on rejected applications, and allows us to imagine the complexity of the process and internal relations.

Existence of credit bureau institutions are the best tools within a country to minimize reject inference bias and help stabilize the banking business and make it more inherently safe.

%% file: abt.tex
\subsection{ABT dataset, variables descriptions \label{opis_abt}}

All variables are described in tables~\ref{abt1}, 
\ref{abt2}, \ref{abt3}, \ref{abt4} i~\ref{abt5}. Only target variables 
are omitted.

\begin{table}
\begin{center}
\caption{ABT variables, part 1}
\label{abt1}
{\scriptsize
\begin{tabular}{|l|l|l|}
 \hline
  & & \\
  Nr & Name & Description \\
  & & \\
 \hline

1 & cid & Id of application \\
2 & aid & Id of Cust. \\
3 & period & Year, month in format YYYYMM \\
4 & act\_age & Actual Cust. age \\
5 & act\_cc & Actual credit capacity  (installment plus spendings) over income \\
6 & act\_loaninc & Loan amount over income \\
7 & app\_income & Cust. income \\
8 & app\_loan\_amount & Loan amount \\
9 & app\_n\_installments & Number of installments \\
10 & app\_number\_of\_children & Number of children \\
11 & app\_spendings & Spendings \\
12 & app\_installment & Installment amount \\
13 & app\_char\_branch & Branch \\
14 & app\_char\_gender & Gender \\
15 & app\_char\_job\_code & Job code \\
16 & app\_char\_marital\_status & Marital status \\
17 & app\_char\_city & City type \\
18 & app\_char\_home\_status & Home status \\
19 & app\_char\_cars & Cars \\
20 & act\_call\_n\_loan & Actual Cust. loan number \\
21 & act\_ccss\_n\_loan & Actual Cust. loan number of Css product \\
22 & act\_cins\_n\_loan & Actual Cust. loan number of Ins product \\
23 & act\_ccss\_maxdue & Cust. actual maximal due installments on product css \\
24 & act\_cins\_maxdue & Cust. actual maximal due installments on product ins \\
25 & act\_ccss\_n\_loans\_act & Cust. actual number of loans on product css \\
26 & act\_cins\_n\_loans\_act & Cust. actual number of loans on product ins \\
27 & act\_ccss\_utl & Cust. actual utilization rate on product css \\
28 & act\_cins\_utl & Cust. actual utilization rate on product ins \\
29 & act\_call\_cc & Cust. credit capacity (all installments plus spendings) over income \\
30 & act\_ccss\_cc & Cust. credit capacity (installment plus spendings) over income on product css \\
31 & act\_cins\_cc & Cust. credit capacity (installment plus spendings) over income on product ins \\
32 & act\_ccss\_dueutl & Cust. due installments over all installments rate on product css \\
33 & act\_cins\_dueutl & Cust. due installments over all installments rate on product ins \\
34 & act\_cus\_active & Cust. had active (status=A) loans one month before \\
35 & act\_ccss\_n\_statB & Cust. historical number of finished loans with status B on product css \\
36 & act\_cins\_n\_statB & Cust. historical number of finished loans with status B on product ins \\
37 & act\_ccss\_n\_statC & Cust. historical number of finished loans with status C on product css \\
38 & act\_cins\_n\_statC & Cust. historical number of finished loans with status C on product ins \\
39 & act\_ccss\_n\_loans\_hist & Cust. historical number of loans on product css \\
40 & act\_cins\_n\_loans\_hist & Cust. historical number of loans on product ins \\

\hline
\end{tabular}
}
\end{center}
\end{table}

\begin{table}
\begin{center}
\caption{ABT variables, part 2}
\label{abt2}
{\scriptsize
\begin{tabular}{|l|l|l|}
 \hline
  & & \\
  Nr & Name & Description \\
  & & \\
 \hline

41 & act\_ccss\_min\_lninst & Cust. minimal number of left installments on product css \\
42 & act\_cins\_min\_lninst & Cust. minimal number of left installments on product ins \\
43 & act\_ccss\_min\_pninst & Cust. minimal number of paid installments on product css \\
44 & act\_cins\_min\_pninst & Cust. minimal number of paid installments on product ins \\
45 & act\_ccss\_min\_seniority & Cust. minimal seniority on product css \\
46 & act\_cins\_min\_seniority & Cust. minimal seniority on product ins \\
47 & act3\_n\_arrears & Cust. number in arrears on all loans during the last 3 months \\
48 & act6\_n\_arrears & Cust. number in arrears on all loans during the last 6 months \\
49 & act9\_n\_arrears & Cust. number in arrears on all loans during the last 9 months \\
50 & act12\_n\_arrears & Cust. number in arrears on all loans during the last 12 months \\
51 & act3\_n\_arrears\_days & Cust. number of days greter than 15 on all loans during the last 3 months \\
52 & act6\_n\_arrears\_days & Cust. number of days greter than 15 on all loans during the last 6 months \\
53 & act9\_n\_arrears\_days & Cust. number of days greter than 15 on all loans during the last 9 months \\
54 & act12\_n\_arrears\_days & Cust. number of days greter than 15 on all loans during the last 12 months \\
55 & act3\_n\_good\_days & Cust. number of days lower than 15 on all loans during the last 3 months \\
56 & act6\_n\_good\_days & Cust. number of days lower than 15 on all loans during the last 6 months \\
57 & act9\_n\_good\_days & Cust. number of days lower than 15 on all loans during the last 9 months \\
58 & act12\_n\_good\_days & Cust. number of days lower than 15 on all loans during the last 12 months \\
59 & act\_ccss\_seniority & Cust. seniority on product css \\
60 & act\_cins\_seniority & Cust. seniority on product ins \\
61 & ags12\_Max\_CMaxC\_Days & Max calc. on last 12 mths on max Cust. days for Css product \\
62 & ags12\_Max\_CMaxI\_Days & Max calc. on last 12 mths on max Cust. days for Ins product \\
63 & ags12\_Max\_CMaxA\_Days & Max calc. on last 12 mths on max Cust. days for all product \\
64 & ags12\_Max\_CMaxC\_Due & Max calc. on last 12 mths on max Cust. due for Css product \\
65 & ags12\_Max\_CMaxI\_Due & Max calc. on last 12 mths on max Cust. due for Ins product \\
66 & ags12\_Max\_CMaxA\_Due & Max calc. on last 12 mths on max Cust. due for all product \\
67 & agr12\_Max\_CMaxC\_Days & Max calc. on last 12 mths on unmissing max Cust. days for Css product \\
68 & agr12\_Max\_CMaxI\_Days & Max calc. on last 12 mths on unmissing max Cust. days for Ins product \\
69 & agr12\_Max\_CMaxA\_Days & Max calc. on last 12 mths on unmissing max Cust. days for all product \\
70 & agr12\_Max\_CMaxC\_Due & Max calc. on last 12 mths on unmissing max Cust. due for Css product \\
71 & agr12\_Max\_CMaxI\_Due & Max calc. on last 12 mths on unmissing max Cust. due for Ins product \\
72 & agr12\_Max\_CMaxA\_Due & Max calc. on last 12 mths on unmissing max Cust. due for all product \\
73 & ags3\_Max\_CMaxC\_Days & Max calc. on last 3 mths on max Cust. days for Css product \\
74 & ags3\_Max\_CMaxI\_Days & Max calc. on last 3 mths on max Cust. days for Ins product \\
75 & ags3\_Max\_CMaxA\_Days & Max calc. on last 3 mths on max Cust. days for all product \\
76 & ags3\_Max\_CMaxC\_Due & Max calc. on last 3 mths on max Cust. due for Css product \\
77 & ags3\_Max\_CMaxI\_Due & Max calc. on last 3 mths on max Cust. due for Ins product \\
78 & ags3\_Max\_CMaxA\_Due & Max calc. on last 3 mths on max Cust. due for all product \\
79 & agr3\_Max\_CMaxC\_Days & Max calc. on last 3 mths on unmissing max Cust. days for Css product \\
80 & agr3\_Max\_CMaxI\_Days & Max calc. on last 3 mths on unmissing max Cust. days for Ins product \\

\hline
\end{tabular}
}
\end{center}
\end{table}

\begin{table}
\begin{center}
\caption{ABT variables, part 3}
\label{abt3}
{\scriptsize
\begin{tabular}{|l|l|l|}
 \hline
  & & \\
   Nr & Name & Description \\
  & & \\
 \hline

81 & agr3\_Max\_CMaxA\_Days & Max calc. on last 3 mths on unmissing max Cust. days for all product \\
82 & agr3\_Max\_CMaxC\_Due & Max calc. on last 3 mths on unmissing max Cust. due for Css product \\
83 & agr3\_Max\_CMaxI\_Due & Max calc. on last 3 mths on unmissing max Cust. due for Ins product \\
84 & agr3\_Max\_CMaxA\_Due & Max calc. on last 3 mths on unmissing max Cust. due for all product \\
85 & ags6\_Max\_CMaxC\_Days & Max calc. on last 6 mths on max Cust. days for Css product \\
86 & ags6\_Max\_CMaxI\_Days & Max calc. on last 6 mths on max Cust. days for Ins product \\
87 & ags6\_Max\_CMaxA\_Days & Max calc. on last 6 mths on max Cust. days for all product \\
88 & ags6\_Max\_CMaxC\_Due & Max calc. on last 6 mths on max Cust. due for Css product \\
89 & ags6\_Max\_CMaxI\_Due & Max calc. on last 6 mths on max Cust. due for Ins product \\
90 & ags6\_Max\_CMaxA\_Due & Max calc. on last 6 mths on max Cust. due for all product \\
91 & agr6\_Max\_CMaxC\_Days & Max calc. on last 6 mths on unmissing max Cust. days for Css product \\
92 & agr6\_Max\_CMaxI\_Days & Max calc. on last 6 mths on unmissing max Cust. days for Ins product \\
93 & agr6\_Max\_CMaxA\_Days & Max calc. on last 6 mths on unmissing max Cust. days for all product \\
94 & agr6\_Max\_CMaxC\_Due & Max calc. on last 6 mths on unmissing max Cust. due for Css product \\
95 & agr6\_Max\_CMaxI\_Due & Max calc. on last 6 mths on unmissing max Cust. due for Ins product \\
96 & agr6\_Max\_CMaxA\_Due & Max calc. on last 6 mths on unmissing max Cust. due for all product \\
97 & ags9\_Max\_CMaxC\_Days & Max calc. on last 9 mths on max Cust. days for Css product \\
98 & ags9\_Max\_CMaxI\_Days & Max calc. on last 9 mths on max Cust. days for Ins product \\
99 & ags9\_Max\_CMaxA\_Days & Max calc. on last 9 mths on max Cust. days for all product \\
100 & ags9\_Max\_CMaxC\_Due & Max calc. on last 9 mths on max Cust. due for Css product \\
101 & ags9\_Max\_CMaxI\_Due & Max calc. on last 9 mths on max Cust. due for Ins product \\
102 & ags9\_Max\_CMaxA\_Due & Max calc. on last 9 mths on max Cust. due for all product \\
103 & agr9\_Max\_CMaxC\_Days & Max calc. on last 9 mths on unmissing max Cust. days for Css product \\
104 & agr9\_Max\_CMaxI\_Days & Max calc. on last 9 mths on unmissing max Cust. days for Ins product \\
105 & agr9\_Max\_CMaxA\_Days & Max calc. on last 9 mths on unmissing max Cust. days for all product \\
106 & agr9\_Max\_CMaxC\_Due & Max calc. on last 9 mths on unmissing max Cust. due for Css product \\
107 & agr9\_Max\_CMaxI\_Due & Max calc. on last 9 mths on unmissing max Cust. due for Ins product \\
108 & agr9\_Max\_CMaxA\_Due & Max calc. on last 9 mths on unmissing max Cust. due for all product \\
109 & ags12\_Mean\_CMaxC\_Days & Mean calc. on last 12 mths on max Cust. days for Css product \\
110 & ags12\_Mean\_CMaxI\_Days & Mean calc. on last 12 mths on max Cust. days for Ins product \\
111 & ags12\_Mean\_CMaxA\_Days & Mean calc. on last 12 mths on max Cust. days for all product \\
112 & ags12\_Mean\_CMaxC\_Due & Mean calc. on last 12 mths on max Cust. due for Css product \\
113 & ags12\_Mean\_CMaxI\_Due & Mean calc. on last 12 mths on max Cust. due for Ins product \\
114 & ags12\_Mean\_CMaxA\_Due & Mean calc. on last 12 mths on max Cust. due for all product \\
115 & agr12\_Mean\_CMaxC\_Days & Mean calc. on last 12 mths on unmissing max Cust. days for Css product \\
116 & agr12\_Mean\_CMaxI\_Days & Mean calc. on last 12 mths on unmissing max Cust. days for Ins product \\
117 & agr12\_Mean\_CMaxA\_Days & Mean calc. on last 12 mths on unmissing max Cust. days for all product \\
118 & agr12\_Mean\_CMaxC\_Due & Mean calc. on last 12 mths on unmissing max Cust. due for Css product \\
119 & agr12\_Mean\_CMaxI\_Due & Mean calc. on last 12 mths on unmissing max Cust. due for Ins product \\
120 & agr12\_Mean\_CMaxA\_Due & Mean calc. on last 12 mths on unmissing
max Cust. due for all product \\
 
\hline
\end{tabular}
}
\end{center}
\end{table}

\begin{table}
\begin{center}
\caption{ABT variables, part 4}
\label{abt4}
{\scriptsize
\begin{tabular}{|l|l|l|}
 \hline
  & & \\
  Nr & Name & Description \\
  & & \\
 \hline

121 & ags3\_Mean\_CMaxC\_Days & Mean calc. on last 3 mths on max Cust. days for Css product \\
122 & ags3\_Mean\_CMaxI\_Days & Mean calc. on last 3 mths on max Cust. days for Ins product \\
123 & ags3\_Mean\_CMaxA\_Days & Mean calc. on last 3 mths on max Cust. days for all product \\
124 & ags3\_Mean\_CMaxC\_Due & Mean calc. on last 3 mths on max Cust. due for Css product \\
125 & ags3\_Mean\_CMaxI\_Due & Mean calc. on last 3 mths on max Cust. due for Ins product \\
126 & ags3\_Mean\_CMaxA\_Due & Mean calc. on last 3 mths on max Cust. due for all product \\
127 & agr3\_Mean\_CMaxC\_Days & Mean calc. on last 3 mths on unmissing max Cust. days for Css product \\
128 & agr3\_Mean\_CMaxI\_Days & Mean calc. on last 3 mths on unmissing max Cust. days for Ins product \\
129 & agr3\_Mean\_CMaxA\_Days & Mean calc. on last 3 mths on unmissing max Cust. days for all product \\
130 & agr3\_Mean\_CMaxC\_Due & Mean calc. on last 3 mths on unmissing max Cust. due for Css product \\
131 & agr3\_Mean\_CMaxI\_Due & Mean calc. on last 3 mths on unmissing max Cust. due for Ins product \\
132 & agr3\_Mean\_CMaxA\_Due & Mean calc. on last 3 mths on unmissing max Cust. due for all product \\
133 & ags6\_Mean\_CMaxC\_Days & Mean calc. on last 6 mths on max Cust. days for Css product \\
134 & ags6\_Mean\_CMaxI\_Days & Mean calc. on last 6 mths on max Cust. days for Ins product \\
135 & ags6\_Mean\_CMaxA\_Days & Mean calc. on last 6 mths on max Cust. days for all product \\
136 & ags6\_Mean\_CMaxC\_Due & Mean calc. on last 6 mths on max Cust. due for Css product \\
137 & ags6\_Mean\_CMaxI\_Due & Mean calc. on last 6 mths on max Cust. due for Ins product \\
138 & ags6\_Mean\_CMaxA\_Due & Mean calc. on last 6 mths on max Cust. due for all product \\
139 & agr6\_Mean\_CMaxC\_Days & Mean calc. on last 6 mths on unmissing max Cust. days for Css product \\
140 & agr6\_Mean\_CMaxI\_Days & Mean calc. on last 6 mths on unmissing max Cust. days for Ins product \\
141 & agr6\_Mean\_CMaxA\_Days & Mean calc. on last 6 mths on unmissing max Cust. days for all product \\
142 & agr6\_Mean\_CMaxC\_Due & Mean calc. on last 6 mths on unmissing max Cust. due for Css product \\
143 & agr6\_Mean\_CMaxI\_Due & Mean calc. on last 6 mths on unmissing max Cust. due for Ins product \\
144 & agr6\_Mean\_CMaxA\_Due & Mean calc. on last 6 mths on unmissing max Cust. due for all product \\
145 & ags9\_Mean\_CMaxC\_Days & Mean calc. on last 9 mths on max Cust. days for Css product \\
146 & ags9\_Mean\_CMaxI\_Days & Mean calc. on last 9 mths on max Cust. days for Ins product \\
147 & ags9\_Mean\_CMaxA\_Days & Mean calc. on last 9 mths on max Cust. days for all product \\
148 & ags9\_Mean\_CMaxC\_Due & Mean calc. on last 9 mths on max Cust. due for Css product \\
149 & ags9\_Mean\_CMaxI\_Due & Mean calc. on last 9 mths on max Cust. due for Ins product \\
150 & ags9\_Mean\_CMaxA\_Due & Mean calc. on last 9 mths on max Cust. due for all product \\
151 & agr9\_Mean\_CMaxC\_Days & Mean calc. on last 9 mths on unmissing max Cust. days for Css product \\
152 & agr9\_Mean\_CMaxI\_Days & Mean calc. on last 9 mths on unmissing max Cust. days for Ins product \\
153 & agr9\_Mean\_CMaxA\_Days & Mean calc. on last 9 mths on unmissing max Cust. days for all product \\
154 & agr9\_Mean\_CMaxC\_Due & Mean calc. on last 9 mths on unmissing max Cust. due for Css product \\
155 & agr9\_Mean\_CMaxI\_Due & Mean calc. on last 9 mths on unmissing max Cust. due for Ins product \\
156 & agr9\_Mean\_CMaxA\_Due & Mean calc. on last 9 mths on unmissing max Cust. due for all product \\
157 & ags12\_Min\_CMaxC\_Days & Min calc. on last 12 mths on max Cust. days for Css product \\
158 & ags12\_Min\_CMaxI\_Days & Min calc. on last 12 mths on max Cust. days for Ins product \\
159 & ags12\_Min\_CMaxA\_Days & Min calc. on last 12 mths on max Cust. days for all product \\
160 & ags12\_Min\_CMaxC\_Due & Min calc. on last 12 mths on max Cust. due for Css product \\

\hline
\end{tabular}
}
\end{center}
\end{table}

\begin{table}
\begin{center}
\caption{ABT variables, part 5}
\label{abt5}
{\scriptsize
\begin{tabular}{|l|l|l|}
 \hline
  & & \\
   Nr & Name & Description \\
  & & \\
 \hline

161 & ags12\_Min\_CMaxI\_Due & Min calc. on last 12 mths on max Cust. due for Ins product \\
162 & ags12\_Min\_CMaxA\_Due & Min calc. on last 12 mths on max Cust. due for all product \\
163 & agr12\_Min\_CMaxC\_Days & Min calc. on last 12 mths on unmissing max Cust. days for Css product \\
164 & agr12\_Min\_CMaxI\_Days & Min calc. on last 12 mths on unmissing max Cust. days for Ins product \\
165 & agr12\_Min\_CMaxA\_Days & Min calc. on last 12 mths on unmissing max Cust. days for all product \\
166 & agr12\_Min\_CMaxC\_Due & Min calc. on last 12 mths on unmissing max Cust. due for Css product \\
167 & agr12\_Min\_CMaxI\_Due & Min calc. on last 12 mths on unmissing max Cust. due for Ins product \\
168 & agr12\_Min\_CMaxA\_Due & Min calc. on last 12 mths on unmissing max Cust. due for all product \\
169 & ags3\_Min\_CMaxC\_Days & Min calc. on last 3 mths on max Cust. days for Css product \\
170 & ags3\_Min\_CMaxI\_Days & Min calc. on last 3 mths on max Cust. days for Ins product \\
171 & ags3\_Min\_CMaxA\_Days & Min calc. on last 3 mths on max Cust. days for all product \\
172 & ags3\_Min\_CMaxC\_Due & Min calc. on last 3 mths on max Cust. due for Css product \\
173 & ags3\_Min\_CMaxI\_Due & Min calc. on last 3 mths on max Cust. due for Ins product \\
174 & ags3\_Min\_CMaxA\_Due & Min calc. on last 3 mths on max Cust. due for all product \\
175 & agr3\_Min\_CMaxC\_Days & Min calc. on last 3 mths on unmissing max Cust. days for Css product \\
176 & agr3\_Min\_CMaxI\_Days & Min calc. on last 3 mths on unmissing max Cust. days for Ins product \\
177 & agr3\_Min\_CMaxA\_Days & Min calc. on last 3 mths on unmissing max Cust. days for all product \\
178 & agr3\_Min\_CMaxC\_Due & Min calc. on last 3 mths on unmissing max Cust. due for Css product \\
179 & agr3\_Min\_CMaxI\_Due & Min calc. on last 3 mths on unmissing max Cust. due for Ins product \\
180 & agr3\_Min\_CMaxA\_Due & Min calc. on last 3 mths on unmissing max Cust. due for all product \\
181 & ags6\_Min\_CMaxC\_Days & Min calc. on last 6 mths on max Cust. days for Css product \\
182 & ags6\_Min\_CMaxI\_Days & Min calc. on last 6 mths on max Cust. days for Ins product \\
183 & ags6\_Min\_CMaxA\_Days & Min calc. on last 6 mths on max Cust. days for all product \\
184 & ags6\_Min\_CMaxC\_Due & Min calc. on last 6 mths on max Cust. due for Css product \\
185 & ags6\_Min\_CMaxI\_Due & Min calc. on last 6 mths on max Cust. due for Ins product \\
186 & ags6\_Min\_CMaxA\_Due & Min calc. on last 6 mths on max Cust. due for all product \\
187 & agr6\_Min\_CMaxC\_Days & Min calc. on last 6 mths on unmissing max Cust. days for Css product \\
188 & agr6\_Min\_CMaxI\_Days & Min calc. on last 6 mths on unmissing max Cust. days for Ins product \\
189 & agr6\_Min\_CMaxA\_Days & Min calc. on last 6 mths on unmissing max Cust. days for all product \\
190 & agr6\_Min\_CMaxC\_Due & Min calc. on last 6 mths on unmissing max Cust. due for Css product \\
191 & agr6\_Min\_CMaxI\_Due & Min calc. on last 6 mths on unmissing max Cust. due for Ins product \\
192 & agr6\_Min\_CMaxA\_Due & Min calc. on last 6 mths on unmissing max Cust. due for all product \\
193 & ags9\_Min\_CMaxC\_Days & Min calc. on last 9 mths on max Cust. days for Css product \\
194 & ags9\_Min\_CMaxI\_Days & Min calc. on last 9 mths on max Cust. days for Ins product \\
195 & ags9\_Min\_CMaxA\_Days & Min calc. on last 9 mths on max Cust. days for all product \\
196 & ags9\_Min\_CMaxC\_Due & Min calc. on last 9 mths on max Cust. due for Css product \\
197 & ags9\_Min\_CMaxI\_Due & Min calc. on last 9 mths on max Cust. due for Ins product \\
198 & ags9\_Min\_CMaxA\_Due & Min calc. on last 9 mths on max Cust. due for all product \\
199 & agr9\_Min\_CMaxC\_Days & Min calc. on last 9 mths on unmissing max Cust. days for Css product \\
200 & agr9\_Min\_CMaxI\_Days & Min calc. on last 9 mths on unmissing max Cust. days for Ins product \\
201 & agr9\_Min\_CMaxA\_Days & Min calc. on last 9 mths on unmissing max Cust. days for all product \\
202 & agr9\_Min\_CMaxC\_Due & Min calc. on last 9 mths on unmissing max Cust. due for Css product \\
203 & agr9\_Min\_CMaxI\_Due & Min calc. on last 9 mths on unmissing max Cust. due for Ins product \\
204 & agr9\_Min\_CMaxA\_Due & Min calc. on last 9 mths on unmissing max Cust. due for all product \\

\hline
\end{tabular}
}
\end{center}
\end{table}

%% file: dokum.tex
\newpage
\subsection{Models documentations \label{dok_modeli}}
\zero
\par

All models are quick and dirty scorecards based on WoE logistic regression approach~\citep{sasbook} made by the author's own SAS 4GL codes.
For every model there a brief of the documentation is presented that consists of: a calibration formula, transforming score value into the probability of the modeling event (based on inverse logit function and regression coefficients calculated by Logistic regression model), Gini values for training and validating datasets, lift statistics (how many times model is better than random one on 1, 5, 10 and 20 percentiles). Finally the scorecard is presented. All partial scores are well calibrated to always have the same value for the worst risky segment (attribute). It results in a simple way to identify the strongest variable in the model, namely the variable with the biggest partial score is the most important in the model.

\subsubsection{Risk PD model for instalment loans (PD Ins)}

Scorecard and some KPIs are presented in table~\ref{modelpdins}. Value PD\_Ins is calibrated as follows:

\texttt{ pd\_ins=1/(1+exp(-(-0.032205144*risk\_ins\_score+9.4025558419))); }

\begin{table}
\begin{center}
\caption{Model PD Ins.}
\label{modelpdins}
\vskip0.5cm
{\scriptsize

\begin{tabular}{ c | c | c | c | c | c }
\hline

Gini (train) & Gini (Valid) & Lift1 & Lift5 & Lift10 & Lift20 \\
\hline
73,37\% & 73,37\% & 7,62 & 5,59 & 4,52 & 3,34 \\

\hline
\end{tabular}

\vskip0.5cm
\centerline{Scorecard}
\begin{tabular}{ l | c | r }
\hline

Variable & Condition & Partial score \\
\hline
ACT\_CC  & 1.0535455861 $<$ ACT\_CC  & -1  \\
   & 0.857442348 $<$ ACT\_CC $\le$ 1.0535455861  & 29  \\
   & 0.3324658426 $<$ ACT\_CC $\le$ 0.857442348  & 40  \\
   & 0.248125937 $<$ ACT\_CC $\le$ 0.3324658426  & 49  \\
   & ACT\_CC $\le$ 0.248125937  & 61  \\
\hline
ACT\_CINS\_MIN\_SENIORITY  & ACT\_CINS\_MIN\_SENIORITY $\le$ 22  & -1  \\
   & 22 $<$ ACT\_CINS\_MIN\_SENIORITY $\le$ 36  & 50  \\
   & missing ACT\_CINS\_MIN\_SENIORITY  & 53  \\
   & 36 $<$ ACT\_CINS\_MIN\_SENIORITY $\le$ 119  & 76  \\
   & 119 $<$ ACT\_CINS\_MIN\_SENIORITY  & 99  \\
\hline
ACT\_CINS\_N\_LOAN  & 1 $<$ ACT\_CINS\_N\_LOAN  & -1  \\
   & ACT\_CINS\_N\_LOAN $\le$ 1  & 57  \\
\hline
ACT\_CINS\_N\_STATC  & ACT\_CINS\_N\_STATC $\le$ 0  & -1  \\
   & 0 $<$ ACT\_CINS\_N\_STATC $\le$ 1  & 49  \\
   & missing ACT\_CINS\_N\_STATC  & 49  \\
   & 1 $<$ ACT\_CINS\_N\_STATC $\le$ 2  & 54  \\
   & 2 $<$ ACT\_CINS\_N\_STATC  & 87  \\
\hline
APP\_CHAR\_JOB\_CODE  & Contract  & -1  \\
   & Owner company  & 58  \\
   & Retired  & 76  \\
   & Permanent  & 81  \\
\hline
APP\_CHAR\_MARITAL\_STATUS  & Single  & -1  \\
   & Divorced  & 40  \\
   & Maried  & 55  \\
   & Widowed  & 57  \\
\hline
APP\_LOAN\_AMOUNT  & 11376 $<$ APP\_LOAN\_AMOUNT  & -1  \\
   & 8880 $<$ APP\_LOAN\_AMOUNT $\le$ 11376  & 21  \\
   & 7656 $<$ APP\_LOAN\_AMOUNT $\le$ 8880  & 30  \\
   & 4824 $<$ APP\_LOAN\_AMOUNT $\le$ 7656  & 35  \\
   & 1920 $<$ APP\_LOAN\_AMOUNT $\le$ 4824  & 51  \\
   & APP\_LOAN\_AMOUNT $\le$ 1920  & 57  \\
\hline
APP\_NUMBER\_OF\_CHILDREN  & APP\_NUMBER\_OF\_CHILDREN $\le$ 0  & -1  \\
   & 0 $<$ APP\_NUMBER\_OF\_CHILDREN $\le$ 1  & 23  \\
   & 1 $<$ APP\_NUMBER\_OF\_CHILDREN  & 57  \\
\hline

\end{tabular}

}
\end{center}
\end{table}

\subsubsection{Risk model PD for cash loans (PD Css)}

Scorecard and some KPIs are presented in table~\ref{modelpdcss}. Value PD\_Css is calibrated as follows:

\texttt{ pd\_css=1/(1+exp(-(-0.028682728*risk\_css\_score+8.1960829753))); }

\begin{table}
\begin{center}
\caption{Model PD Css.}
\label{modelpdcss}
\vskip0.5cm
{\scriptsize

\begin{tabular}{ c | c | c | c | c | c }
\hline
Gini (train) & Gini (Valid) & Lift1 & Lift5 & Lift10 & Lift20 \\
\hline
74,06\% & 74,06\% & 1,77 & 1,67 & 1,63 & 1,64 \\

\hline
\end{tabular}

\vskip0.5cm
\centerline{Scorecard}
\begin{tabular}{ l | c | r }
\hline
Variable & Condition & Partial score \\
\hline
ACT\_AGE  & 50 $<$ ACT\_AGE $\le$ 61  & 24  \\
   & 62 $<$ ACT\_AGE $\le$ 68  & 34  \\
   & ACT\_AGE $\le$ 50  & 33  \\
   & 68 $<$ ACT\_AGE $\le$ 80  & 44  \\
   & 61 $<$ ACT\_AGE $\le$ 62  & 46  \\
   & 80 $<$ ACT\_AGE  & 70  \\
\hline
ACT\_CALL\_CC  & 1.5775700935 $<$ ACT\_CALL\_CC $\le$ 2.0091145833  & 24  \\
   & 2.0091145833 $<$ ACT\_CALL\_CC  & 34  \\
   & 1.4502074689 $<$ ACT\_CALL\_CC $\le$ 1.5775700935  & 43  \\
   & 1.1900674433 $<$ ACT\_CALL\_CC $\le$ 1.4502074689  & 51  \\
   & ACT\_CALL\_CC $\le$ 1.1900674433  & 64  \\
\hline
ACT\_CCSS\_DUEUTL  & 0.0416666667 $<$ ACT\_CCSS\_DUEUTL $\le$ 0.21875  & 24  \\
   & 0.21875 $<$ ACT\_CCSS\_DUEUTL  & 29  \\
   & 0.025 $<$ ACT\_CCSS\_DUEUTL $\le$ 0.0416666667  & 33  \\
   & 0.0208333333 $<$ ACT\_CCSS\_DUEUTL $\le$ 0.025  & 41  \\
   & ACT\_CCSS\_DUEUTL $\le$ 0.0208333333  & 54  \\
   & missing ACT\_CCSS\_DUEUTL  & 57  \\
\hline
ACT\_CCSS\_MIN\_LNINST  & 1 $<$ ACT\_CCSS\_MIN\_LNINST $\le$ 7  & 24  \\
   & 7 $<$ ACT\_CCSS\_MIN\_LNINST $\le$ 11  & 26  \\
   & ACT\_CCSS\_MIN\_LNINST $\le$ 0  & 31  \\
   & 11 $<$ ACT\_CCSS\_MIN\_LNINST  & 32  \\
   & 0 $<$ ACT\_CCSS\_MIN\_LNINST $\le$ 1  & 40  \\
   & missing ACT\_CCSS\_MIN\_LNINST  & 47  \\
\hline
ACT\_CCSS\_N\_STATC  & 0 $<$ ACT\_CCSS\_N\_STATC $\le$ 4  & 24  \\
   & ACT\_CCSS\_N\_STATC $\le$ 0  & 32  \\
   & 4 $<$ ACT\_CCSS\_N\_STATC $\le$ 10  & 34  \\
   & 10 $<$ ACT\_CCSS\_N\_STATC $\le$ 21  & 51  \\
   & missing ACT\_CCSS\_N\_STATC  & 53  \\
   & 21 $<$ ACT\_CCSS\_N\_STATC  & 82  \\
\hline
AGS3\_MEAN\_CMAXA\_DUE  & 1.333 $<$ AGS3\_MEAN\_CMAXA\_DUE  & 24  \\
   & 0.666 $<$ AGS3\_MEAN\_CMAXA\_DUE $\le$ 1.333  & 47  \\
   & 0.333 $<$ AGS3\_MEAN\_CMAXA\_DUE $\le$ 0.666  & 62  \\
   & AGS3\_MEAN\_CMAXA\_DUE $\le$ 0.333  & 73  \\
\hline
APP\_NUMBER\_OF\_CHILDREN  & APP\_NUMBER\_OF\_CHILDREN $\le$ 0  & 24  \\
   & 0 $<$ APP\_NUMBER\_OF\_CHILDREN $\le$ 1  & 33  \\
   & 1 $<$ APP\_NUMBER\_OF\_CHILDREN  & 57  \\
\hline

\end{tabular}

}
\end{center}
\end{table}

\subsubsection{Risk model PD for future cash loans at the time of
instalment application (Cross PD Css)}

\texttt{ cross\_pd\_css=1/(1+exp(-(-0.028954669*cross\_css\_score+8.2497434934))); }

\begin{table}
\begin{center}
\caption{Model Cross PD Css.}
\label{modelcrosspdcss}
\vskip0.5cm
{\scriptsize

\begin{tabular}{ c | c | c | c | c | c }
\hline
Gini (train) & Gini (Valid) & Lift1 & Lift5 & Lift10 & Lift20 \\
\hline
73,77\% & 73,77\% & 1,80 & 1,50 & 1,52 & 1,48 \\

\hline
\end{tabular}

\vskip0.5cm
\centerline{Scorecard}
\begin{tabular}{ l | c | r }
\hline
Variable & Condition & Partial score \\
\hline
ACT12\_N\_GOOD\_DAYS  & 4 $<$ ACT12\_N\_GOOD\_DAYS $\le$ 8  & 29  \\
   & 3 $<$ ACT12\_N\_GOOD\_DAYS $\le$ 4  & 34  \\
   & 8 $<$ ACT12\_N\_GOOD\_DAYS  & 34  \\
   & 2 $<$ ACT12\_N\_GOOD\_DAYS $\le$ 3  & 37  \\
   & ACT12\_N\_GOOD\_DAYS $\le$ 2  & 45  \\
   & missing ACT12\_N\_GOOD\_DAYS  & 54  \\
\hline
ACT\_CCSS\_MAXDUE  & 1 $<$ ACT\_CCSS\_MAXDUE $\le$ 4  & 29  \\
   & 4 $<$ ACT\_CCSS\_MAXDUE  & 37  \\
   & 0 $<$ ACT\_CCSS\_MAXDUE $\le$ 1  & 45  \\
   & ACT\_CCSS\_MAXDUE $\le$ 0  & 59  \\
   & missing ACT\_CCSS\_MAXDUE  & 71  \\
\hline
ACT\_CCSS\_N\_STATC  & ACT\_CCSS\_N\_STATC $\le$ 4  & 29  \\
   & 4 $<$ ACT\_CCSS\_N\_STATC $\le$ 6  & 40  \\
   & 6 $<$ ACT\_CCSS\_N\_STATC $\le$ 15  & 54  \\
   & missing ACT\_CCSS\_N\_STATC  & 79  \\
   & 15 $<$ ACT\_CCSS\_N\_STATC $\le$ 26  & 85  \\
   & 26 $<$ ACT\_CCSS\_N\_STATC  & 125  \\
\hline
ACT\_CCSS\_UTL  & ACT\_CCSS\_UTL $\le$ 0.4083333333  & 29  \\
   & 0.4083333333 $<$ ACT\_CCSS\_UTL $\le$ 0.4479166667  & 32  \\
   & 0.4479166667 $<$ ACT\_CCSS\_UTL $\le$ 0.4895833333  & 36  \\
   & 0.4895833333 $<$ ACT\_CCSS\_UTL $\le$ 0.5208333333  & 40  \\
   & 0.5208333333 $<$ ACT\_CCSS\_UTL $\le$ 0.5347222222  & 44  \\
   & missing ACT\_CCSS\_UTL  & 57  \\
   & 0.5347222222 $<$ ACT\_CCSS\_UTL  & 58  \\
\hline
AGS3\_MEAN\_CMAXA\_DUE  & 1 $<$ AGS3\_MEAN\_CMAXA\_DUE $\le$ 3  & 29  \\
   & 3 $<$ AGS3\_MEAN\_CMAXA\_DUE  & 33  \\
   & 0.6666666667 $<$ AGS3\_MEAN\_CMAXA\_DUE $\le$ 1  & 39  \\
   & AGS3\_MEAN\_CMAXA\_DUE $\le$ 0.6666666667  & 49  \\
   & missing AGS3\_MEAN\_CMAXA\_DUE  & 60  \\
\hline
APP\_INCOME  & APP\_INCOME $\le$ 411  & 29  \\
   & 411 $<$ APP\_INCOME $\le$ 573  & 42  \\
   & 3872 $<$ APP\_INCOME  & 52  \\
   & 573 $<$ APP\_INCOME $\le$ 1049  & 60  \\
   & 1049 $<$ APP\_INCOME $\le$ 3872  & 77  \\
\hline

\end{tabular}

}
\end{center}
\end{table}

\subsubsection{Propensity model, probability of future cash response (PR) at the time of first credit application (PR Css)}

\texttt{ pr\_css=1/(1+exp(-(-0.035007455*response\_score+10.492092793))); }

\begin{table}
\begin{center}
\caption{Model PR Css.}
\label{modelprcss}
\vskip0.5cm
{\scriptsize

\begin{tabular}{ c | c | c | c | c | c }
\hline

Gini (train) & Gini (Valid) & Lift1 & Lift5 & Lift10 & Lift20 \\
\hline
86,22\% & 86,22\% & 4,36 & 3,03 & 2,79 & 2,60 \\

\hline
\end{tabular}

\vskip0.5cm
\centerline{Scorecard}
\begin{tabular}{ l | c | r }
\hline
Variable & Condition & Partial score \\
\hline
ACT\_CCSS\_MIN\_SENIORITY  & ACT\_CCSS\_MIN\_SENIORITY $\le$ 4  & 51  \\
   & 4 $<$ ACT\_CCSS\_MIN\_SENIORITY $\le$ 7  & 60  \\
   & 30 $<$ ACT\_CCSS\_MIN\_SENIORITY  & 64  \\
   & 7 $<$ ACT\_CCSS\_MIN\_SENIORITY $\le$ 9  & 65  \\
   & 9 $<$ ACT\_CCSS\_MIN\_SENIORITY $\le$ 30  & 76  \\
   & missing ACT\_CCSS\_MIN\_SENIORITY  & 96  \\
\hline
ACT\_CCSS\_N\_LOAN  & 5 $<$ ACT\_CCSS\_N\_LOAN  & 51  \\
   & 4 $<$ ACT\_CCSS\_N\_LOAN $\le$ 5  & 59  \\
   & 2 $<$ ACT\_CCSS\_N\_LOAN $\le$ 4  & 80  \\
   & 1 $<$ ACT\_CCSS\_N\_LOAN $\le$ 2  & 108  \\
   & ACT\_CCSS\_N\_LOAN $\le$ 0  & 124  \\
   & 0 $<$ ACT\_CCSS\_N\_LOAN $\le$ 1  & 131  \\
\hline
ACT\_CCSS\_N\_STATC  & 12 $<$ ACT\_CCSS\_N\_STATC  & 51  \\
   & 7 $<$ ACT\_CCSS\_N\_STATC $\le$ 12  & 60  \\
   & 3 $<$ ACT\_CCSS\_N\_STATC $\le$ 7  & 68  \\
   & ACT\_CCSS\_N\_STATC $\le$ 3  & 78  \\
   & missing ACT\_CCSS\_N\_STATC  & 106  \\
\hline
ACT\_CINS\_N\_STATC  & 5 $<$ ACT\_CINS\_N\_STATC  & 51  \\
   & 3 $<$ ACT\_CINS\_N\_STATC $\le$ 5  & 54  \\
   & 1 $<$ ACT\_CINS\_N\_STATC $\le$ 3  & 59  \\
   & 0 $<$ ACT\_CINS\_N\_STATC $\le$ 1  & 64  \\
   & ACT\_CINS\_N\_STATC $\le$ 0  & 75  \\
   & missing ACT\_CINS\_N\_STATC  & 103  \\
\hline

\end{tabular}

}
\end{center}
\end{table}

\section{Some additional files}
The example of SAS dataset calculated for the first strategy and more detailed strategy reports can be found on the web adress:

\centerline{\texttt{http://kprzan.w.interia.pl/ThePowerOfCreditScoring.rar}}